\documentclass[iop,apj,tighten,numberedappendix,twocolappendix,revtex4,letterpaper]{emulateapj}

\usepackage[pdfa,breaklinks,colorlinks,urlcolor=blue,citecolor=blue,linkcolor=blue]{hyperref}
\usepackage{natbib}
\usepackage{apjfonts}
\usepackage{amsmath}
\usepackage{graphicx}
\usepackage{bm}
\usepackage{paralist}
\usepackage[capitalize,noabbrev]{cleveref}
\usepackage{color}

\graphicspath{{figures/}}

\usepackage{etoolbox}
\makeatletter
\patchcmd{\NAT@citex}
  {\@citea\NAT@hyper@{%
     \NAT@nmfmt{\NAT@nm}%
     \hyper@natlinkbreak{\NAT@aysep\NAT@spacechar}{\@citeb\@extra@b@citeb}%
     \NAT@date}}
  {\@citea\NAT@nmfmt{\NAT@nm}%
   \NAT@aysep\NAT@spacechar\NAT@hyper@{\NAT@date}}{}{}
\patchcmd{\NAT@citex}
  {\@citea\NAT@hyper@{%
     \NAT@nmfmt{\NAT@nm}%
     \hyper@natlinkbreak{\NAT@spacechar\NAT@@open\if*#1*\else#1\NAT@spacechar\fi}%
       {\@citeb\@extra@b@citeb}%
     \NAT@date}}
  {\@citea\NAT@nmfmt{\NAT@nm}%
   \NAT@spacechar\NAT@@open\if*#1*\else#1\NAT@spacechar\fi\NAT@hyper@{\NAT@date}}
  {}{}
\makeatother

\usepackage{xspace}

\newcommand{\mean}[1]{\ensuremath{\langle#1\rangle}}

\newcommand{\cps}{\,s$^{-1}$\xspace}

\submitted{Received 2013 March 29; accepted 2013 June 3; published 2013 July 26}
\journalinfo{The Astrophysical Journal, 773:66 (12pp), 2013 August 10}

\begin{document}

\title{On Detecting Transient Phenomena}
\shorttitle{On Detecting Transient Phenomena}

\author{G.\ B\'elanger}
\affil{European Space Astronomy Centre (ESA/ESAC), Science Operations Department,  Villanueva de la Ca\~nada (Madrid), Spain;  \href{mailto:gbelanger@sciops.esa.int}{gbelanger@sciops.esa.int}}

\shortauthors{B\'elanger}

\begin{abstract}
Transient phenomena are interesting and potentially highly revealing of details about the processes under observation and study that could otherwise go unnoticed. It is therefore important to maximize the sensitivity of the method used to identify such events. In this article we present a general procedure based on the use of the likelihood function for identifying transients that is particularly suited for real-time applications, because it requires no grouping or pre-processing of the data. The method makes use of all the information that is available in the data throughout the statistical decision making process, and is suitable for a wide range of applications. Here we consider those most common in astrophysics which involve searching for transient sources, events or features in images, time series, energy spectra, and power spectra, and demonstrate the use of the method in the case of a weak X-ray flare in a time series and a short-lived quasi-periodic oscillation in a power spectrum. We derive a fit statistic that is ideal for fitting arbitrarily shaped models to a power density distribution, which is of general interest in all applications involving periodogram analysis.
\end{abstract}

\section{Introduction}
\label{s:introduction}

Some physical processes can be considered continuous in the sense that the discretization of measurements comes from the detector's sampling rate. Others are discrete in the sense that they give rise to individual events that can be detected as such, as long as the sampling is faster than the detection rate. For non-variable processes, the values of the measurements will generally be distributed as either a normal variable---in those seen as continuous, or as a Poisson variable---in those that are discrete. 

The way in which the measurements are distributed is absolutely crucial in applying the appropriate statistical treatment. But irrespective of that distribution, each measurement carries information that can be used to estimate the values of the parameters of models that help us learn about the processes being observed. Most importantly, each measurement considered individually, and the collection of measurements as a whole, all carry statistical evidence that can be used to accurately assess the agreement between a given hypothesis or model and the data.

Treating data as evidence is a powerful means to detect changes, differences, deviations or variations in any kind of process that is observed. Treating data as statistical evidence is, in fact, the only way that data should be treated no matter what the application or circumstances, because that is what data actually is. The way this is done, mathematically and statistically, is through the likelihood function.

As obvious as this is, it is important to point out that the detection of an event or feature that is localized in time, involves identifying something that was not there before. Whether it rises, dwells, and decays over weeks and months like a supernova, or whether it just appears and disappears in a fraction of a second like a $\gamma$-ray burst; whether it manifests as a complete change of shape of the energy spectrum during a state transition in a black hole, or as the short-lived emission line from an accretion event; whether it comes as a sudden change of spectral index in the power spectrum or as the appearance of an ephemeral quasi-periodic oscillation (QPO); all of these phenomena, independently of their particular timescale, share in common that they appear as a sharp change in the data.

Detection and identification of a transient feature in an ensemble of measurements is a statistical procedure that involves comparing numbers and making decisions based on probabilities or, in fact, on probability ratios, and in other words, on likelihoods. Naturally, we would like to use the most powerful method for the problem at hand. Thus, whatever the application, we want to maximize sensitivity to transient phenomena, and minimize the frequency of identifying a statistical fluctuation as an actual event. For this reason we must use all the information that is available, without degrading in any way the data the instruments provide us with, and interpret these as statistical evidence.

In this article we address this task of identifying \emph{transients}, in the most general sense of the word, without reference to the kind of phenomenon nor the kind of data we are working with. In light of this, we use the word \emph{transient} in a sense that is broader than its customary usage, which refers to a source whose intensity varies dramatically enough to go from being invisible to detectable or even bright, and back to being undetectable. We define a \emph{transient} as any feature or change that can be identified in the data as \emph{statistically distinct} from the global process and underlying conditions. This definition therefore implies that if a feature cannot be distinguished by statistical means, it cannot be detected and identified. Whether this is because the transient is too weak or too long-lived does not matter. The limitations of a particular detection procedure, with its own thresholds and timescales, can always be accurately established before applying it.

We develop and present a general method based on the likelihood function that can be applied to a wide range of problems (Section\;\ref{s:identifyingTransients}). We describe the procedure (Section\;\ref{s:method}), what the likelihood function is (Section\;\ref{s:likelihoodFunction}), and illustrate the method for a general counting experiment (Section\;\ref{s:illustrationOfMethod}). We elaborate on the procedure's use and applicability in astronomy and astrophysics (Section\;\ref{s:astrophysics}) by considering, after a few introductory remarks (Section\;\ref{s:astroIntro}), its application to images (Section\;\ref{s:images}), time series (Section\;\ref{s:timeSeries}), energy spectra (Section\;\ref{s:energySpectra}), and power spectra (Section\;\ref{s:powerSpectra}). We briefly discuss identification of transients over a variable background in the next section (Section\;\ref{s:identifyingTransients}) and in the concluding remarks (Section\;\ref{s:conclusion}). 

The mathematical statistics of likelihood are from the work of \cite{Fisher:1912uv,1922RSPTA.222..309F}; the philosophical basis for, understanding of, and inspiration to work with and interpret data as statistical evidence are primarily from \cite{Royall:1997vc}; and other technical details of data analysis and statistics are mostly from \cite{1997sda..book.....C}.

\section{Identifying Transients}
\label{s:identifyingTransients}

A transient can only be identified as such in relation to something else: in relation to the underlying  background conditions. There are two families of circumstances pertaining to the characteristics of the background process that cover all cases under which transients may appear: the process can be constant or it can be variable. In absolute terms, it could be said that if a process is not constant, then it is variable. In practice, however, the variability is characterized by timescales above which the process is seen to be variable, but below which it cannot, or at least not easily be seen to be variable. Furthermore, the variability will in general manifest differently at different timescales. 

In most applications where transient sources are searched for, they are not detectable until they brighten for a while before once more fading away below the detection level. Therefore, the background is characterized by the statistical and instrumental fluctuations inherent to the particular experimental setup and sky pixel where the source appears. This is also generally true for supernovae and $\gamma$-ray bursts at all wavelengths (but on different timescales), as well as for X-ray novae or X-ray flares, bursts or flashes in most systems: the underlying background is usually constant or very weakly variable in comparison to the sharp change in intensity associated with the transient phenomenon.\footnote{The work of \cite{1998ApJ...504..405S} and \cite{2013ApJ...764..167S} on astronomical time series, with which we were not acquainted while developing our method, is different in its details but quite similar in spirit (even if more complicated in presentation and implementation) and seems well suited for $\gamma$-ray burst searches in X-ray and $\gamma$-rays time series.} In the energy and power spectra, irrespective of spectral shape, transient phenomena will also most often appear against a constant or very weakly variable background. Therefore, in all these cases and in the majority of applications searching for transients, the background is constant or nearly so. 

In fact, this is indeed what allows these short-lived phenomena to be considered and labeled as transient. However, as the intensity of the background against which we seek to identify a transient of a given magnitude increases, the ability to do so decreases. If instead of increasing in intensity the background were to increase in its variability, the ability to identify a transient similarly decreases. In both cases, it is a question of scales: of the intensity scale and of the timescale. In the first case, statistical identification depends mostly on the ratio of intensities between the transient and the background (the commonly held notion of signal-to-noise ratio), but also on its duration: the brighter, the easier to identify; and if it is not bright, the longer it lasts, the easier to identify. In the second, identification depends more on the ratio of the transient's duration to the timescale of the variability of the background, but obviously also on the magnitude of its intensity: the shorter in duration, the easier to identify with respect to a slowly variable background; and naturally the brighter it is, the easier the identification.

Hence, these factors both come into play, and gain or lose in importance depending on the characteristics of the physical process being observed, and also, most sensitively, on the quality of the data. It is important to underline that these elements---intensity ratios and variability timescales---can and should be considered and worked with as parameters in order to establish optimal detection algorithms and thresholds for different problems in various kinds of data sets, as well as to determine the limitations of our methods that should ultimately always be statistical in nature, and dependent on the quality and kind of data, not on weaknesses in the methods themselves. We will carry out and present such an investigation and quantitative characterization of these issues, including the discussion of commonly encountered complications in a future work, together with applications to data from different experiments and surveys.

The purpose of this paper is to present the foundational aspects of the method and illustrate how it can be of immediate applicability in a variety of situations that include the search for transient features in X-ray and $\gamma$-ray data, transient sources in optical sky surveys, radio transients, and in particular rotating radio transients that are currently typically identified by eye. Our basic working assumption, which does indeed cover most applications, is that we are dealing with a background that is constant (nil or not) on the timescales relevant to the problem of identifying transients. 

The method is straight forward and based on the likelihood function. As such, all of the details that pertain to the inherent statistical properties of the kind of random variable that results from the observational process (e.g., normal, Poisson, exponential) are automatically and effortlessly taken into account at every step, and integrated in every aspect of the procedure. The presentation is intended to be as clear, intuitive and easy as possible, with minimal formalism. It is worth noting that the tools for this have been available in practice at least since \cite{1922RSPTA.222..309F}, and in principle since \cite{Bayes:1763ee}. The method makes use of every measurement, does not require grouping or approximations, and therefore does not impose on the data any degradation of its resolution or accuracy, no matter what kind of data it is. Since the approach is the same in all circumstances where the aim is to detect a transient, it is described in general terms and its principles demonstrated in the next section. Astrophysics applications are discussed afterward.

\subsection{Methodology\label{s:method}}

The first measurement gives the first estimate of the reference value: the value we expect to measure under usual conditions when there is no transient. With this single measurement we can already draw the curve that expresses the likelihood of all possible reference values given the evidence carried by that measurement.\footnote{In practice, most observations (from most patches of sky and most sources) are not the first of their kind, and there is, therefore, no need to make a guess of the expected intensity; it can just be determined from previous observations, which implies that even the first measurement can be compared to the expected reference value, and the sensitivity of the method does not depend on the number of prior measurements.} To draw this curve, we must make an informed assumption as to \emph{how the measurements will be distributed} around the true value: we must make an informed assumption about the nature of that random variable. As was mentioned, the most common in physical experiments are the normal distribution seen in continuous processes like measuring temperature, and the Poisson distribution that arises when discrete events are recorded. The likelihood function shows the maximum likelihood of the true value, and the ratio between that and every other possible value: it gives us the relative likelihood of any value with respect to any other, depending only on the nature of the variable and on the data.

The second measurement gives a second estimate of the reference value. Because we already have a likelihood function based on the first measurement, the value of the second can be immediately evaluated for its potential of being a transient---a feature that stands out from what is expected---by computing its likelihood ratio with respect to the previously estimated maximum likelihood. Although limited by the accuracy with which the mean is known, this is nonetheless the only mathematically correct way to evaluate the likelihood of measuring that second value in light of the first, without invoking an a priori model or assumption. If the second measurement is not statistically different from the first beyond the established threshold, it is combined with the first to better estimate the mean, and can also be used to compute a first estimate of the standard deviation of the distribution. The joint likelihood function is computed from the two measurements and its central peak immediately begins to grow narrower and hone in on the true value of the parameter.

The third measurement gives a third estimate of the reference, the likelihood of measuring such a value is evaluated by the ratio of the single-measurement likelihood function centered on the maximum likelihood reference value given by the previously calculated joint likelihood. This is an important point: the joint likelihood is the likelihood function of the reference value given the entire set of measurements considered together as a whole, and with each additional measurement, it gets narrower and more finely peaked on the maximum likelihood reference value; the single-measurement likelihood is the function that shows how likely it is to measure any given value each time a measurement is made. The more precisely the reference value is determined, the more reliable the location of the single-measurement likelihood function. However, its shape depends only on the probability density of the random variable and on the reference value.\footnote{We can formally incorporate the uncertainty in the determination of the reference value into the single-measurement likelihood function by computing the cross-correlation of the joint and single-measurement functions, an operation which yields a broadened function that tends to the pure probability density as the joint likelihood tends toward a delta function. We have investigated this, and found that it increases the complexity of the procedure substantially, but that the effect is only appreciable in the first few measurements where a broadening is seen. Beyond even a handful of measurements, the joint likelihood function is narrow enough for the broadening to be negligible. It is therefore not warranted to include this step unless we are dealing with a process that will only ever yield a handful of measurements.}

With each subsequent measurement, the same procedure is repeated:
\begin{inparaenum}[(1)]
\item compute the likelihood of the newly measured value based on the single-measurement function defined by the current maximum likelihood reference value;
\item if the likelihood is less than the defined threshold, issue a transient event trigger. Do not update the estimate of the reference value;
\item if the likelihood is more than the defined threshold (within the likelihood interval), recalculate the joint likelihood function including the new measurement and update the maximum likelihood reference value.
\end{inparaenum}
The single or multiple thresholds used to identify the occurrence of a transient event must be defined and optimized according to the application.

\begin{figure*}[t]
\epsscale{1.0}
\begin{center}
\includegraphics[scale=0.22, angle=-90]{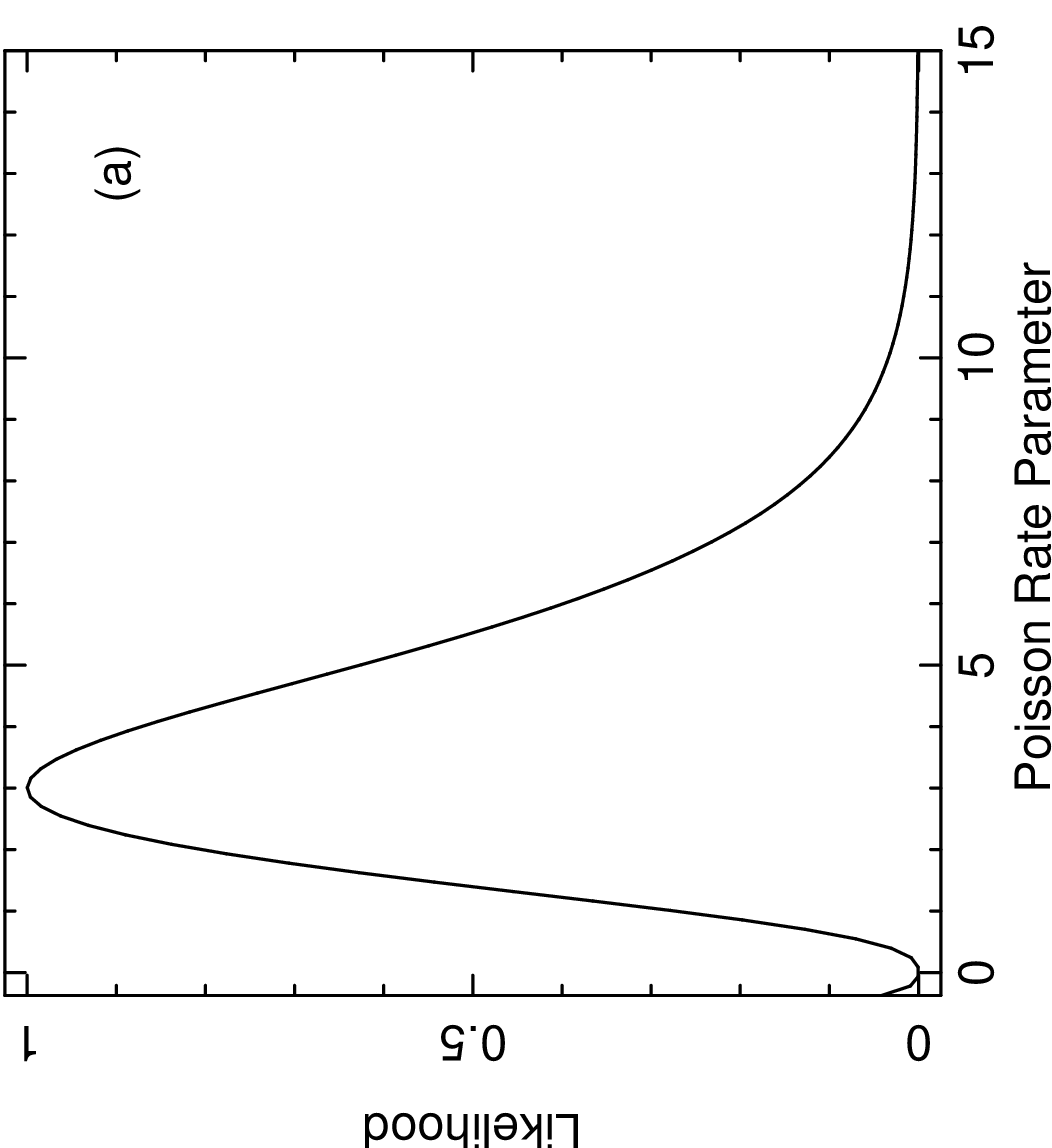}
\includegraphics[scale=0.22, angle=-90]{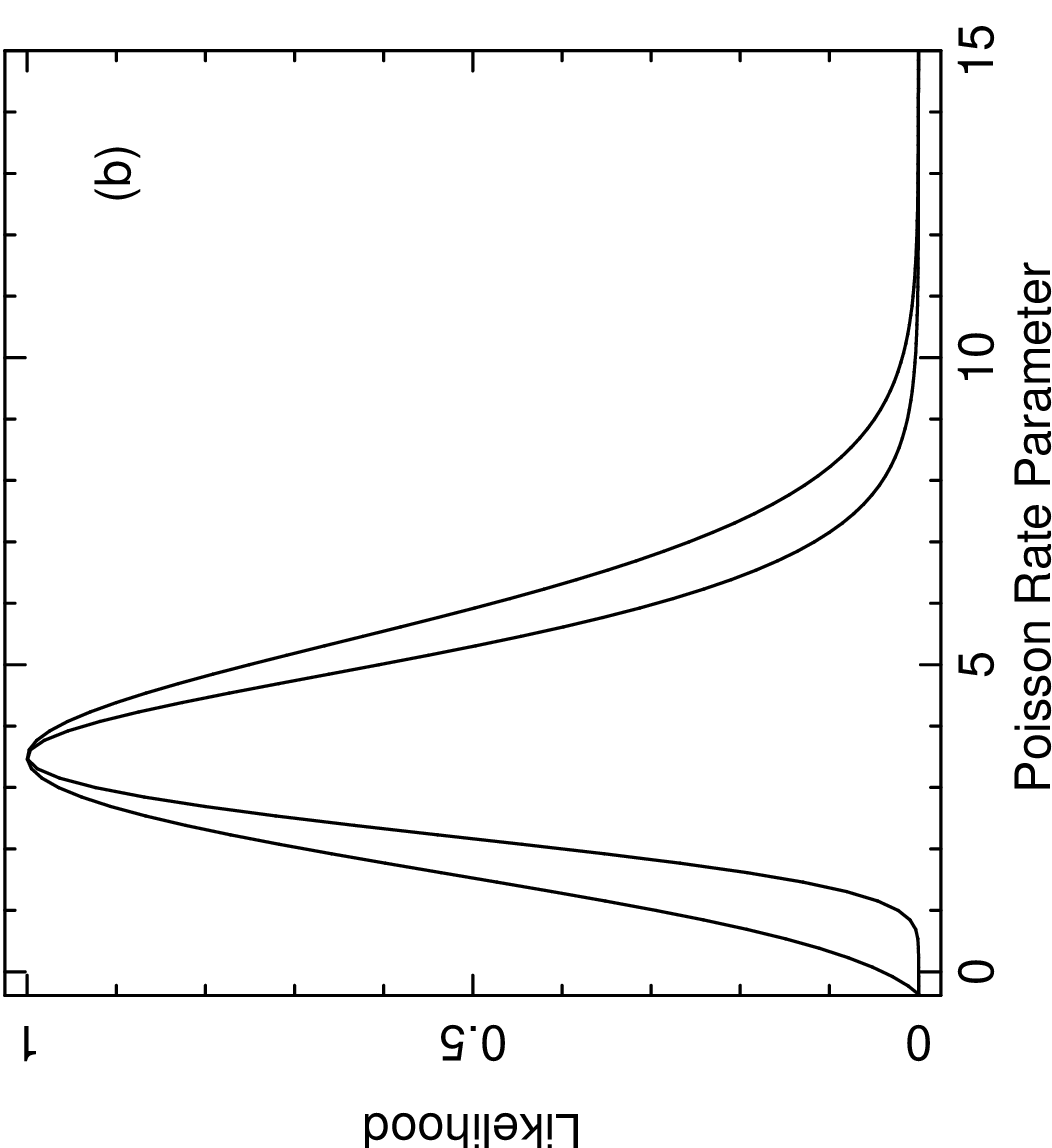}
\includegraphics[scale=0.22, angle=-90]{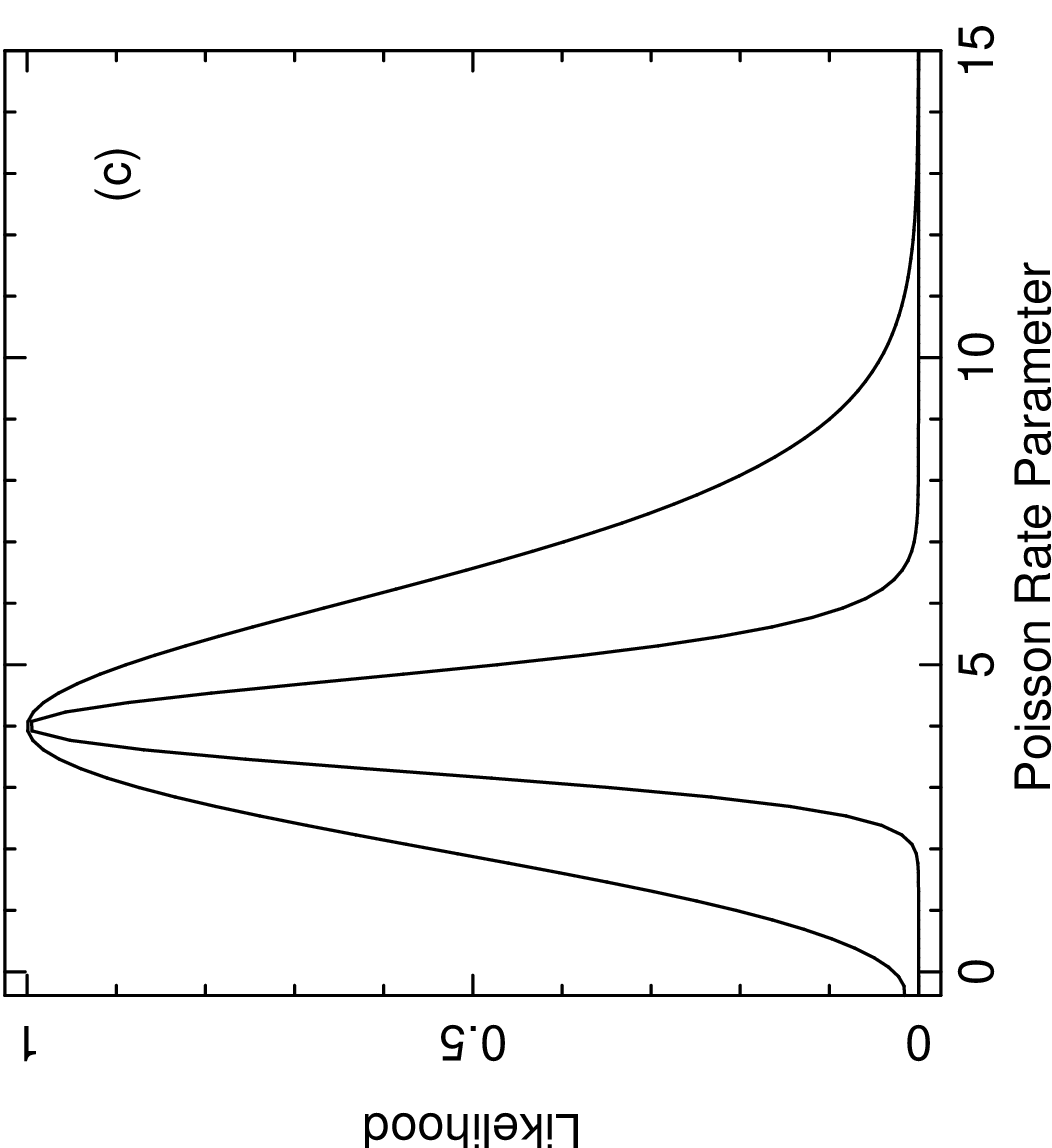}
\includegraphics[scale=0.22, angle=-90]{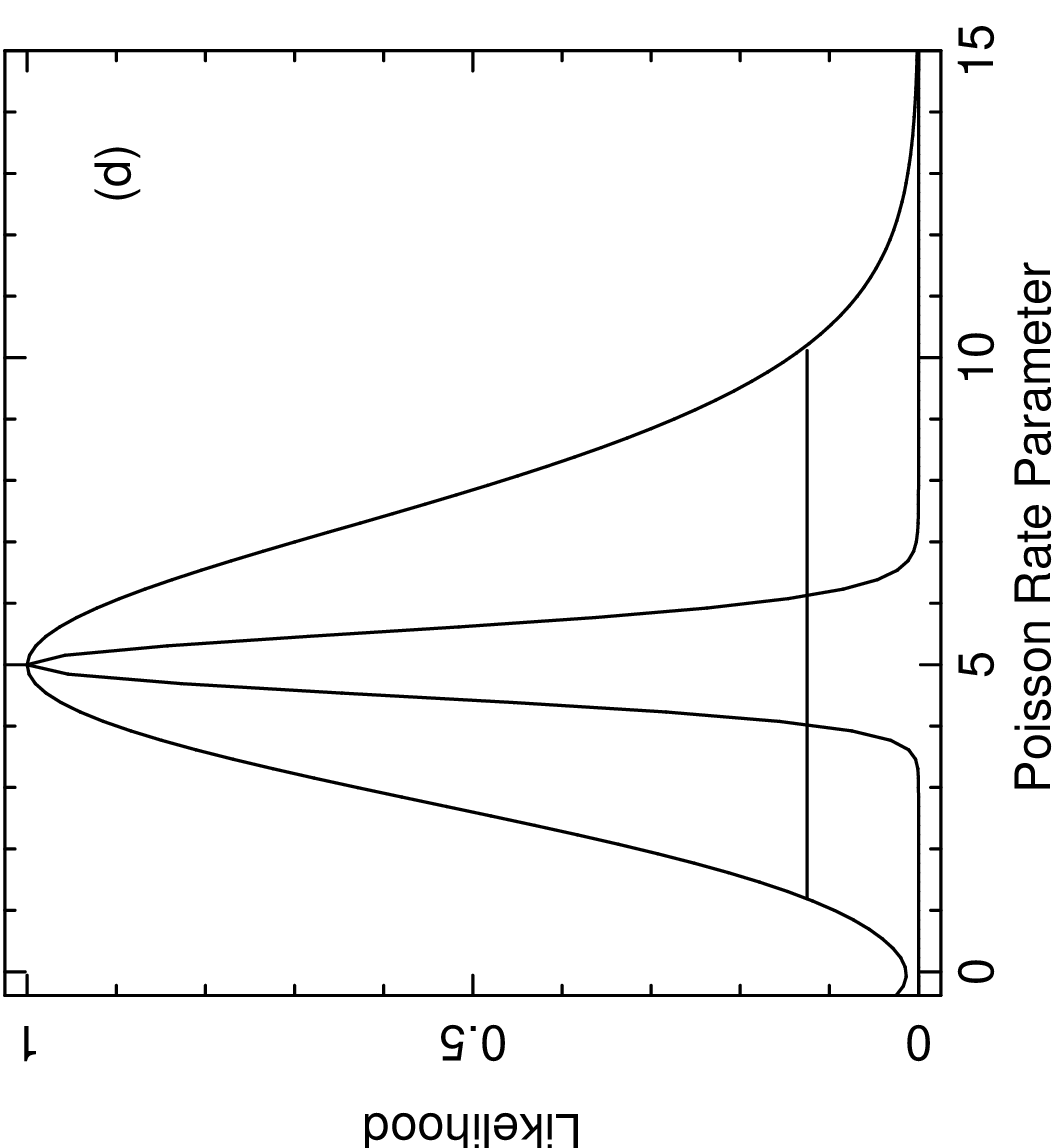}
\end{center}
\caption{\footnotesize Graphical representation of the single-measurement and joint likelihood functions after one (panel (a)), two (panel (b)), seven (panel (c)) and nineteen (panel (d)) measurements of a Poisson variable with a true rate parameter of $\nu=5$. The 1/8 likelihood interval is shown in panel (d).
\label{f:illustration}}
\end{figure*}

\subsection{The Likelihood Function}
\label{s:likelihoodFunction}
The likelihood is proportional to the product of probabilities. However, because there are an infinite number of points on the probability density function, the product of a collection of probabilities will in general be unbounded. Hence, only the \emph{ratio} of likelihoods is meaningful. Although it is reminiscent of a probability distribution, it is quite distinct, because only its shape matters while its scale does not. In the words of \citet[][p.\ 327]{1922RSPTA.222..309F}, ``the likelihood may be held to measure the degree of our rational belief in a conclusion'', and in those of \citet[][p.\ 28]{Royall:1997vc}, ``Probabilities measure uncertainty and likelihood ratios measure evidence.'' It is worth emphasizing the following point: before making a measurement, we have probabilities; after making the measurement, we have likelihoods. We adopt the convenient approach suggested by \cite{1922RSPTA.222..309F} himself, and normalize the likelihood function such that the maximum equals one. This then implies that the value of every point on the curve is already the ratio to the maximum likelihood.

The most fundamental distinction is that probability density relates to the random variable whereas likelihood relates to the parameter. The probability density function expresses how we can expect the measured values of the random variable to be distributed given a certain value of the parameter. For example, if we take the rate parameter $\nu$ of the Poisson distribution to be equal to 5 events per second, the density function tells us that the probability to detect 5 events in a one second interval is given by the value of the density function at 5 and equals 14.6\%, or that the probability to detect 10 events or more is given by the sum from 10 onward and equals 1.4\%. 

Now, knowing that we are observing a Poisson process, say that we measure five events in the sampling time of one second. That measurement is made and the uncertainty about its value therefore disappears. The value is now evidence about the probability distribution, about the unknown value of the parameter. The likelihood function represents this evidence: it tells us that the most likely value of the rate parameter $\nu$ is 5, and that, for example, it is 1.1 times more likely than 6, 4.6 times more likely than 10, and 32 times more likely than 13.4. Computing the likelihood function, the $y$-axis is the relative likelihood and the $x$-axis represents different values of the parameter. In the case of a single measurement $n_0$, the likelihood associated with each value of the parameter $\nu$ on the abscissa is given by (proportional to) $f(n_0;\nu)$; for two measurements $n_1$ and $n_2$, it is given by the product $f(n_1;\nu) f(n_2; \nu)$.

In this work, we consider five families of random variables: the Poisson, normal, $\chi^2$, exponential, and inverse-exponential. For simplicity and clarity of presentation, the relevant equations are given in the Appendix instead of being presented or derived in the text.

\subsection{Illustration of Method}
\label{s:illustrationOfMethod}

The instrument is turned on and the observation begins. Nothing is known other than that we are observing a non-variable Poisson process. The sampling rate is one second and in the first interval three events are detected. With this first measurement, we can already compute the likelihood function of the rate parameter $\nu$, and because we have only one measurement, the joint likelihood is identical to the single-measurement likelihood (\cref{f:illustration}, panel (a)). 

In the second interval, four events are detected. We calculate the likelihood ratio for this measurement in relation to the previously estimated maximum likelihood value of the rate parameter which was 3, and find that it is $L_1(\nu=4)/L_1(\nu=3) = 0.872$ which is much larger than the warning threshold of $1/8 = 0.125$. Therefore, we compute the joint likelihood function using both measurements, and now have a slightly more accurate estimate of the rate parameter as exhibited by the narrower peak of the joint likelihood; the single-measurement function is also updated accordingly (\cref{f:illustration}, panel (b)).

The observation continues and by the time we have made 7 measurements the joint likelihood is quite a bit narrower (\cref{f:illustration}, panel (c)), and after 19 measurements the peak is significantly narrower and peaks on the true rate parameter, $\nu_{t}=5$ (\cref{f:illustration}, panel (d)). Naturally, the more measurements are made, the sharper the peak grows, and hence the accuracy of our maximum likelihood estimate of the rate parameter which in turn defines the single-measurement likelihood function against which we evaluate the evidence for the occurrence of a transient with each new measurement by calculating the likelihood ratio; the 1/8 likelihood interval is drawn and seen to extend from 1.2 to 10.2 in panel (d).

\section{Astrophysical Transients}
\label{s:astrophysics}

\subsection{Introductory Remarks}
\label{s:astroIntro}

The bounds between which changes can be detected are defined by the instrument's sampling rate for the shortest timescales, and by the time spanned by the data on the longest: if changes in the system take place much faster than the sampling rate (millisecond pulses sampled on second timescales), or much slower than the duration of the observation (an outburst that lasts longer than the time during which it was observed without appreciable change in brightness), they will not be readily detectable. Naturally, transient searches only have meaning within these bounds.

Within these limits, the granularity in time of each iteration of the likelihood procedure is a crucial factor. If made fine enough compared to the timescale of the transient, the change will be gradual and smooth enough to be unrecognized as such. Therefore, the time between iterations must be chosen to ensure optimal sensitivity to timescales relevant to the problem. If there are several, the solution is to run multiple procedures in parallel, each with a different characteristic timescale. Another solution is to monitor the evolution of the average value of the likelihood ratio as a function of time. This technique relies on the stability of the value of the likelihood ratio, and is as sensitive as our knowledge of the background conditions against which we want to detect transient events (which grows with each additional measurement), and most importantly on the probability distribution of the measurements. Naturally, we can look at the evolution of the likelihood ratio in each channel, in the joint function or in both to maximize our sensitivity to specific kinds of transients.

\subsection{Transients in Images}
\label{s:images}

Imaging data is acquired by different means at different wavelengths, but in what concerns the task of identifying transient features in these images, the main requirement is that there must be at least more than two and preferably a collection of images of the same region of the sky. Thus, independently of the actual integration time of each snapshot or the means by which this integration time is chosen, we can treat the problem of having sky pixels, each with an ensemble of measured values that we take as an intensity regardless of the units. 

In regards to the intensity, there are two cases: \begin{inparaenum}[(1)] \item the intensity in every pixel is independent of (not correlated with) the intensity in any other pixel, or \item the intensity in a given pixel is related to the intensity of neighboring pixels. \end{inparaenum} If the instrument's point spread function (PSF) is mostly contained within a single detector pixel, then we consider each pixel to give independent intensity measurements and also define the size of independent sky pixels. If this is not the case and the PSF spreads on several detector pixels, then we can either make sky pixels as large as necessary to include a large enough fraction of the PSF in order to be considered independent of neighboring sky pixels, or we must include the model of the PSF in the analysis. 

If pixel intensities are independent of one another, this makes an image a collection of intensity measurements, one per pixel, that each corresponds to a distinct region of the sky. Each snapshot yields one such intensity measurement for each pixel, and therefore the problem immediately reduces to the illustrative case above: with the first snapshot, we already have the means to construct the model for each pixel of what can be expected; with each additional snapshot, the model improves in accuracy, and consequently, our ability to detect changes in intensity increases; and the expected intensity in each sky pixel is represented by the single-measurement likelihood function centered and scaled in accord with the joint likelihood function constructed from the ensemble of intensity measurements. 

The two relevant functional forms are those of the Poisson and the normal density distributions (\cref{eq:poissonDensity,eq:normalDensity}). The procedure was illustrated in Section\;\ref{s:illustrationOfMethod} using the Poisson function (\cref{eq:poissonLikelihood-single,eq:poissonLikelihood-joint}), but maybe in most imaging applications, the normal likelihood function (\cref{eq:normalLikelihood-single,eq:normalLikelihood-joint}) will be the appropriate one to use.

\begin{figure*}
\epsscale{1.0}
\begin{center}
\includegraphics[scale=0.34, angle=-90]{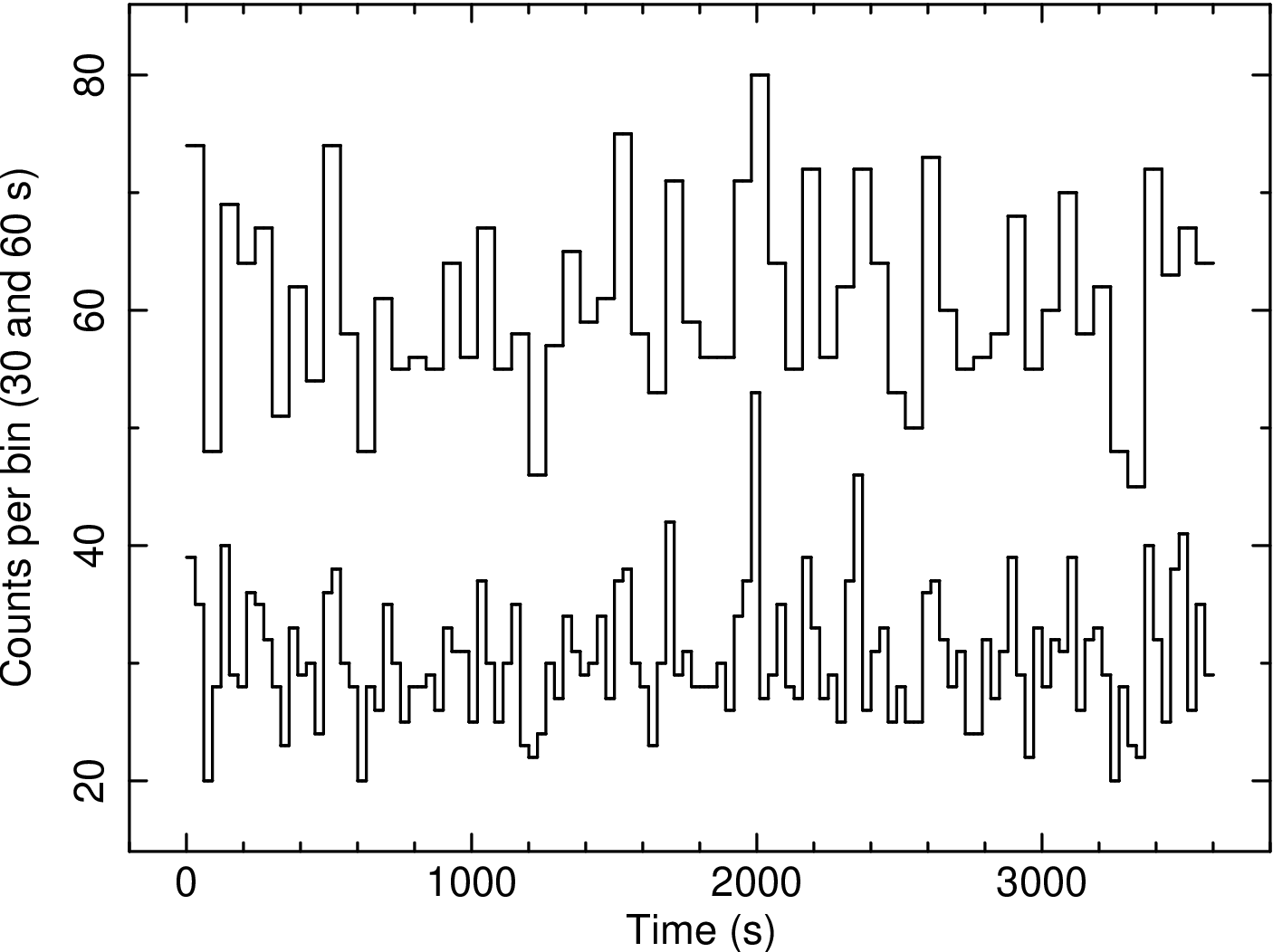} \hspace{2mm}
\includegraphics[scale=0.34, angle=-90]{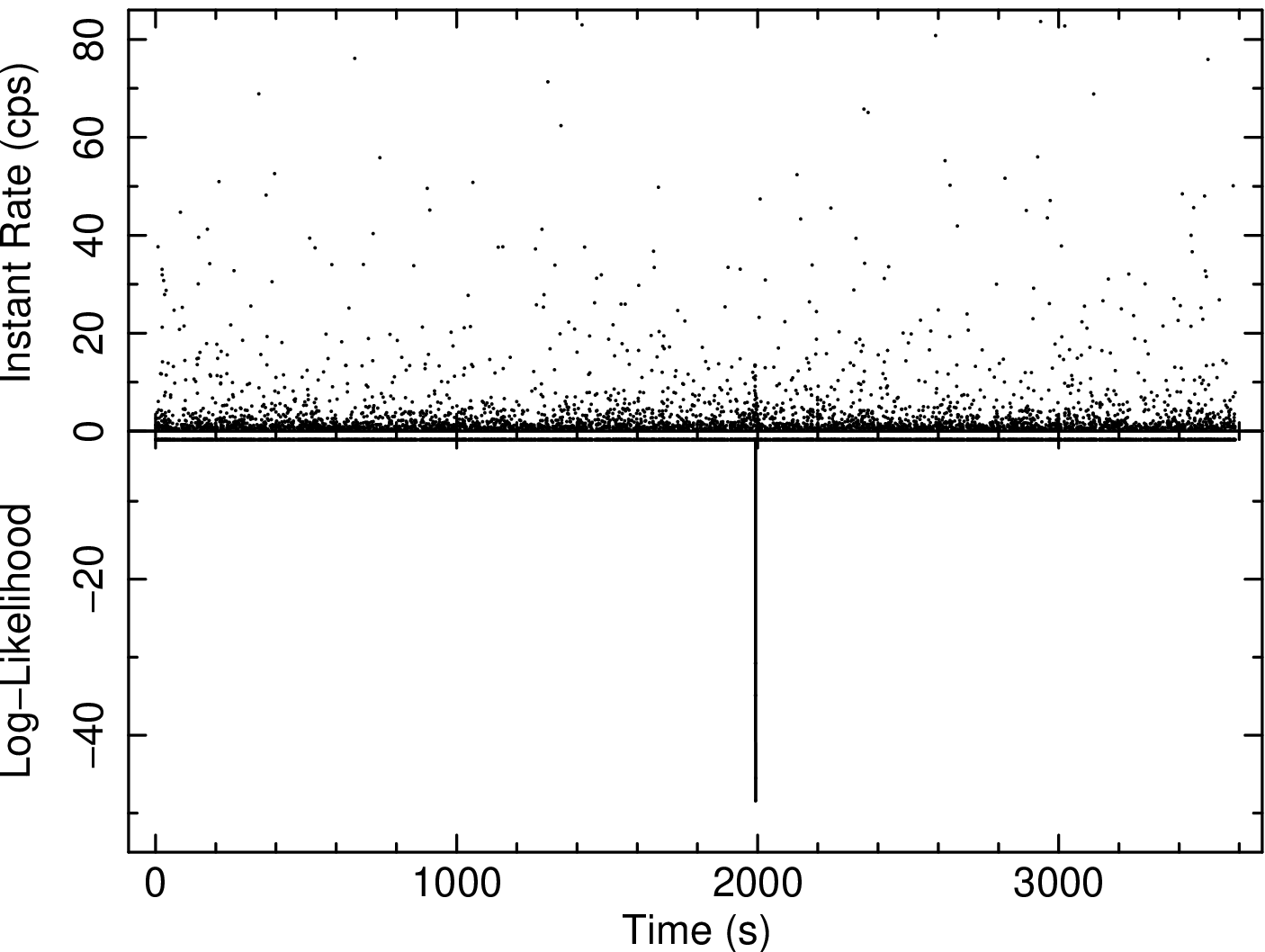}
\end{center}
\caption{\footnotesize Time series of the simulated observation shown in counts per bin for 30 and 60\,s bins (left panel, bottom and top respectively),  and instantaneous count rate calculated as the inverse of the time between events shown as a function of each event's arrival time with the transient detection likelihood also evaluated in real time (right panel, top and bottom respectively). The maximum value of the instantaneous rate is 2950\cps, but the scale is truncated to 86\cps to match the scale of the 60\,s time series and better show the scatter. Values of the log-likelihood that do not meet the trigger criterion are shown at the warning threshold level of $-2.1$ (likelihood of 0.14). The sole detection is that of the transient event, and it dips down to $-48.41$ (likelihood of $10^{-21}$). 
\label{f:transient}}
\end{figure*}

If the detector pixels are smaller than the PSF and the intensity measurements are hence correlated with neighboring ones, the problem is more complex in its details but the same in its principle. The most significant difference in the approach is that instead of treating images as a collection of independent intensity measurements,  they must be considered and modeled as a whole. The global model will preferably be constructed before the observation to include all known sources in the field. With the first image, an intensity is estimated for each identified source included in the model taking into account the instrument's PSF, and then monitored using each subsequent snapshot. Accordingly, each source has its own joint likelihood function that is better defined with each additional image, as well as its own single-measurement likelihood function based on the joint likelihood. These likelihood functions, however, are based on the model of the PSF, and thus only indirectly on the analytical probability density.

In addition, to the iteratively refined modeling of likelihood functions for each identified source, every other pixel in the image is monitored as if it were independent of all others in the same way as in the previous case. This is obviously of prime importance given that our aim is to detect transients and especially new, weak, and invisible or otherwise unidentified sources. When the intensity of one of these pixels is seen to climb or fall out of its likelihood interval in a number of consecutive snapshots, the new source is then properly modeled using the PSF as for all other sources in the field. The detailed treatment depends on the PSF and is not carried out here for a hypothetical case, but the use of a global model is analogous to the treatment of both energy and power spectra. We thus leave out of this section a more explicit discussion.

\subsection{Transients in Time Series}
\label{s:timeSeries}

The procedure for treating time series is, in fact, identical to the one for a single independent sky pixel. And here also, no grouping or resampling of any kind is necessary such that all the information is used with full resolution to yield maximum accuracy in our estimates. 

If the intensity measurements are derived from snapshots, as is often the case at many wavelengths, then a time series is made up of intensities from a single pixel in the images as a function of time. If instead, measurements consist of the detection of individual photons, then this can be treated either as in the general illustration of the method in Section\;\ref{s:illustrationOfMethod}, or, to use every event as each one is detected, the rate parameter can be estimated directly by the inverse of the time between events. For example, a quarter of a second between two events gives a value of the estimate of the intensity in events per second (the rate parameter) of 4\cps; if five seconds pass, then it is 0.2\cps. Therefore, whether the intensity is measured in a sky pixel as a function of time in successive images, whether it is the number of events detected during a given time interval, or whether it is estimated by the inverse of the inter-arrival times for each and every detected photon, the procedure remains the same in all respects and is as in Section\;\ref{s:illustrationOfMethod}. There is, however, a crucial difference in these three closely related cases that must be highlighted. 

In the first case, we can expect the intensity values to be either Poisson or normally distributed depending on the imaging instrument and characteristics of the experiment; in the second, they will always follow the Poisson distribution; and in the third, because inter-arrival times are exponentially distributed, the corresponding \emph{instantaneous rates will follow the inverse-exponential distribution}. The exact distribution of a given random variable is the key to both the accuracy and the power of the likelihood function in statistical analysis.

Figure \ref{f:transient} is a demonstration of the technique applied to an unbinned time series (a list of event arrival times). In this example, we are observing a hypothetical flaring X-ray source embedded in a region from which the average event rate is 1\cps. The observation lasted one hour, and contained a weak flare that lasted about 30\,s and counted exactly 33 events from start to finish. Even though a hint of its presence is seen in the 30\,s binned time series (but not with 60\,s bins), this event could easily have gone unnoticed if the detection relied on the examination of similar time series. 

With the procedure described above, the flare is detected at a log-likelihood of -48.41, and thus likelihood of about $10^{-21}$. This is the combined likelihood of detecting at least eight consecutive events, each of which met the warning threshold, when we expect the mean detection rate. In contrast, looking at the peak that stands out in the 30\,s binned time series, we would compare a total intensity of 53 events against the expectation of 30, and find a likelihood of $5.8\times 10^{-4}$, which might be enough to make us raise an eyebrow, but not much more. This also helps illustrate the important difference between the binned treatment of the data and the unbinned likelihood approach that treats each measurement individually.

In this example with a weak transient, the warning threshold was set at the relatively high log-likelihood value of $-2.1$ (likelihood of 0.12), but the detection trigger was set to require eight consecutive events over the warning threshold.\footnote{This detection strategy was optimized using simulations: when the observation did not include a burst, false positives were detected only 10\% of the time, but when a flare was included, although its strength is indeed very weak, it was detected about 60\% of the time.} Similar searches will require this kind of strategy. For detecting strong but  short-lived transients, it would instead be better to use a very low, single-point threshold. Each application has its own purpose and thus optimal settings that can be very precisely tuned with simulations. 

Related considerations will be discussed in detail in a subsequent paper, but it is worth noting again the significant difference between using the approach described here compared to that of binning the time series and looking for outliers. Although to establish the optimal warning threshold and the number of consecutive warnings required to define a transient is indeed akin to defining, by whatever means, the optimal bin time for a particular kind of transient (e.g., \cite{1998ApJ...504..405S}), it is very different in several ways. Using each measurement individually gives access to the full resolution of the data, and thus the exact distribution in time of the measurements. This, in turn, allows us to legitimately ask the question for each measurement `what is the likelihood of this measurement if we expect that?' (the single-measurement likelihood), and also to ask the question `what is the likelihood of these two, three, four or however many measurements taken together as an ensemble, when we expect this value?' (the joint likelihood), and the answer is always perfectly well-defined and directly interpretable as statistical evidence with respect to the question we are asking, (the hypothesis we are testing). Using each measurement also allows us to discriminate between consecutive measured values above the warning threshold, and an concurrence of several separate fluctuations above that threshold that are not consecutive but happened to occur in the same bin time interval.

\subsection{Transients in Energy Spectra}
\label{s:energySpectra}

In energy spectra, frequency histograms as a function of energy, transient features can also appear. When they do, they can potentially give powerful clues as to the physical changes in the conditions of the system under study. As with images, there are two ways of approaching the problem: each energy channel can be treated as independent of the others, in which case the procedure is the same as that for independent pixels, or we can recognize that the intensity in each spectral channel is almost surely related to the intensity in every other channel because radiative processes, other than those that manifest in mono-energetic line emission, produce a continuum of radiation. The former is simple, relies on no additional information and can thus be very powerful to detect narrow features. It is, however, limited in its sensitivity to changes that occur in several channels simultaneously. The latter is more complex and relies on the use of a global spectral model in order to treat the problem with all channels considered together, but is substantially more sensitive to subtle changes.

Hence, in the absence of a model, we use the former approach, treating each energy channel individually. We emphasize the importance of using all spectral information provided by the instrument and not grouping channels into coarser energy bins because \emph{every measurement counts}, and each one must be used and accounted for. 

Thus, with the very first measurement that falls in a single energy channel, the procedure is initiated and we have an estimate of the expected value that is used to construct the first likelihood function for that channel; with the second in the same channel, the likelihood ratio of that measured value with respect to the previous is calculated and evaluated for its potential as signaling a transient event, following which the joint and single-measurement likelihood functions are updated; with the third and every subsequent measurement, the same procedure is followed.

With a reliable model, even if it is simple, the power to detect changes in the shape of the spectrum increases markedly because the amount of information contained within and used to iteratively update the likelihood function is greater: all the measurements in each channel add to our knowledge of every other channel and makes the overall likelihood function as informative as it can be. The means by which it is calculated and updated is slightly different. In whichever way the measured values are distributed in a channel, it is the model that links these across the entire spectral range, and the key component from which the likelihood function is constructed and on which the entire procedure hinges. 

From an initial estimate of the model's parameter values we define the joint likelihood function. In the case of Poisson distributed data, the function is the product of the contributing elements from each channel, each supplying a term of the form given in Equation (\ref{eq:poissonDensity}): $$f(n;\nu) = \frac{\nu^n e^{-\nu}}{n!}.$$
This yields an ensemble of such terms, each with a different value of $n$: $n_i$---the measured intensity in channel $i$, and a different value of $\nu$: $\nu_i$---the model-predicted intensity in that channel. There is thus a single likelihood function: the joint likelihood function for the model across all spectral channels, and it is given by Equation (\ref{eq:poissonLikelihood-model}): $$L(\pmb{\nu}|\pmb{n}) = \prod_i \frac{{\nu_i}^{n_i} e^{-\nu_i}}{n_i!}.$$ 

\begin{figure*}[htb]
\epsscale{1.0}
\begin{center}
\includegraphics[scale=0.225, angle=-90]{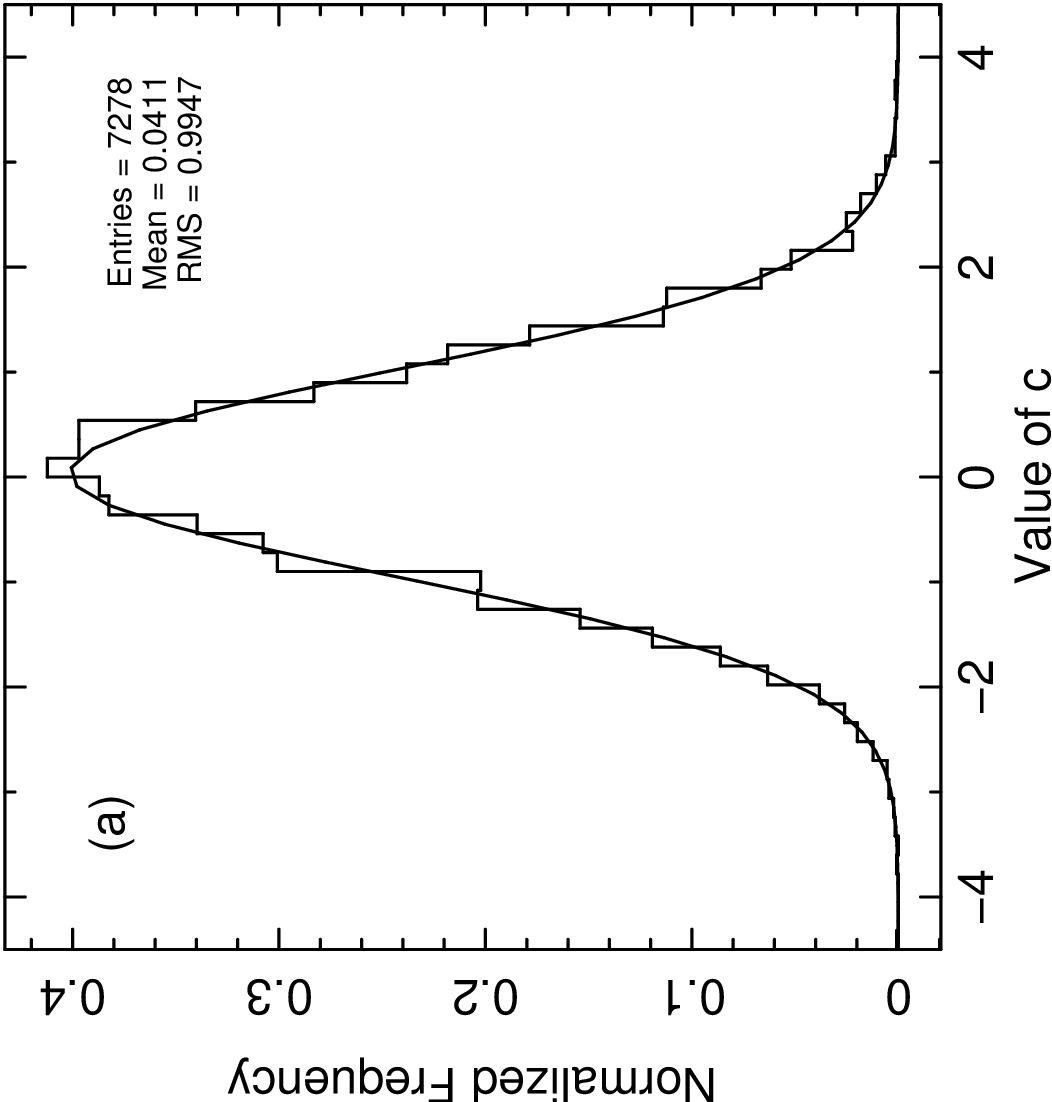}
\includegraphics[scale=0.225, angle=-90]{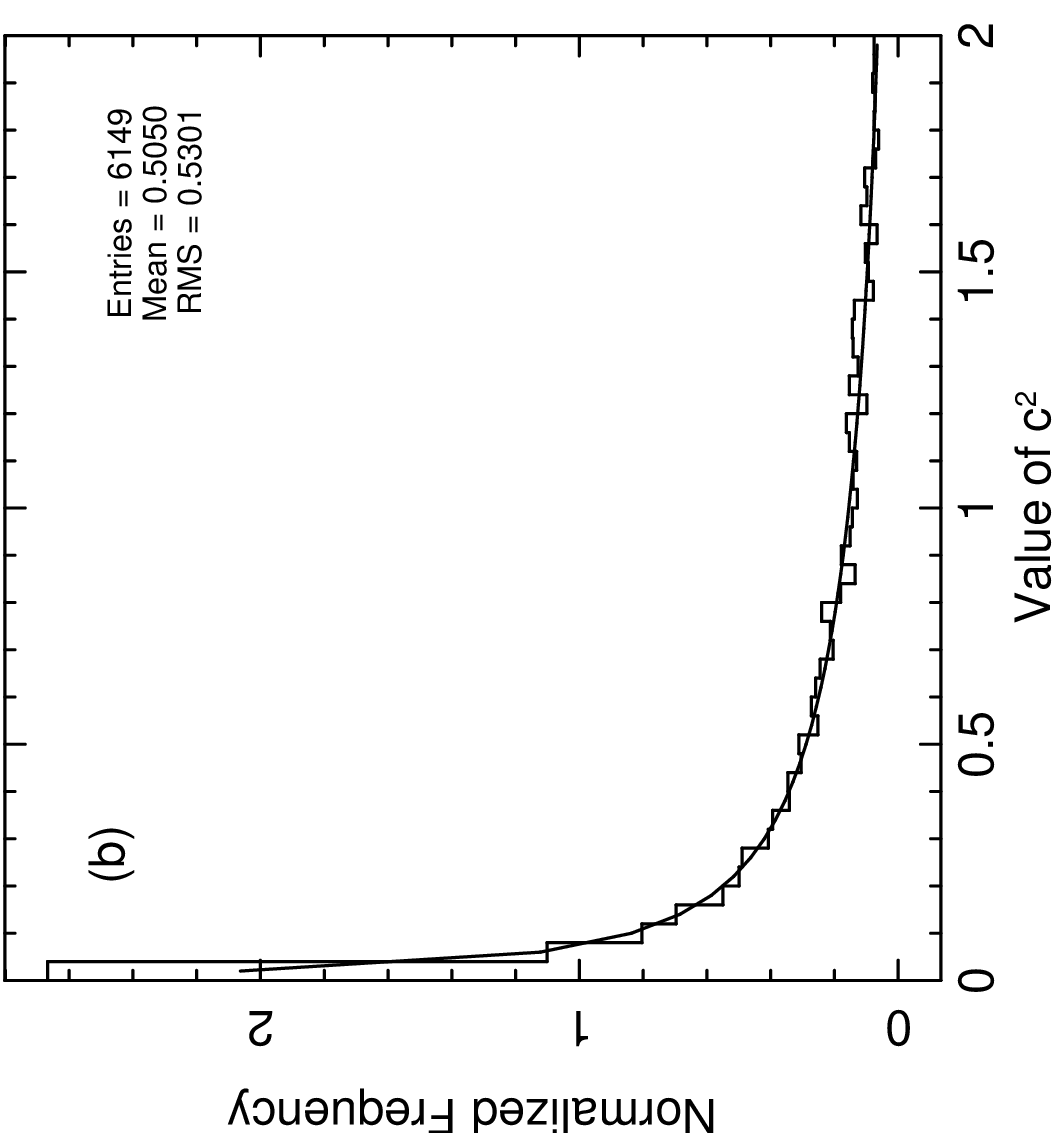}
\includegraphics[scale=0.225, angle=-90]{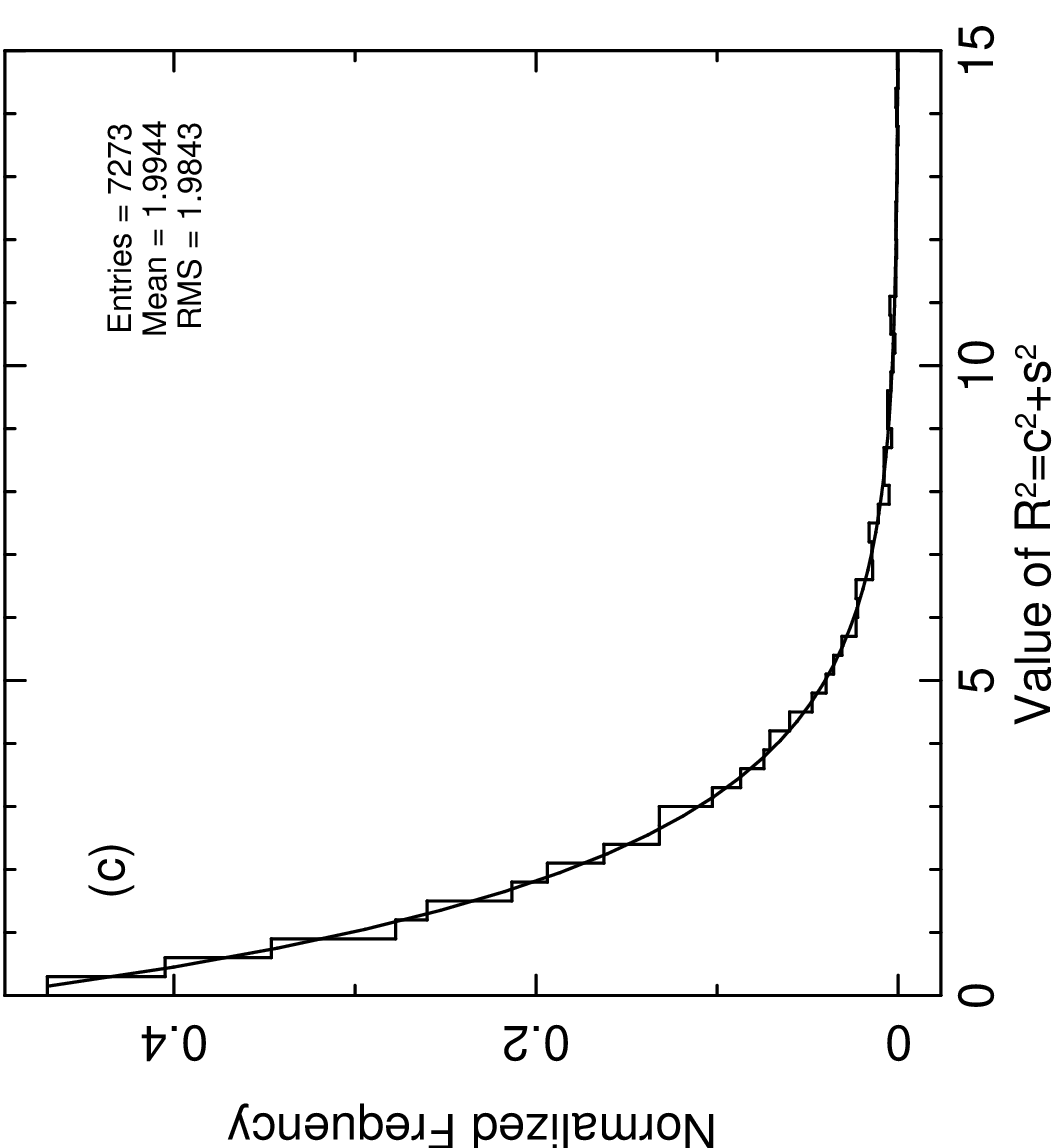}
\includegraphics[scale=0.225, angle=-90]{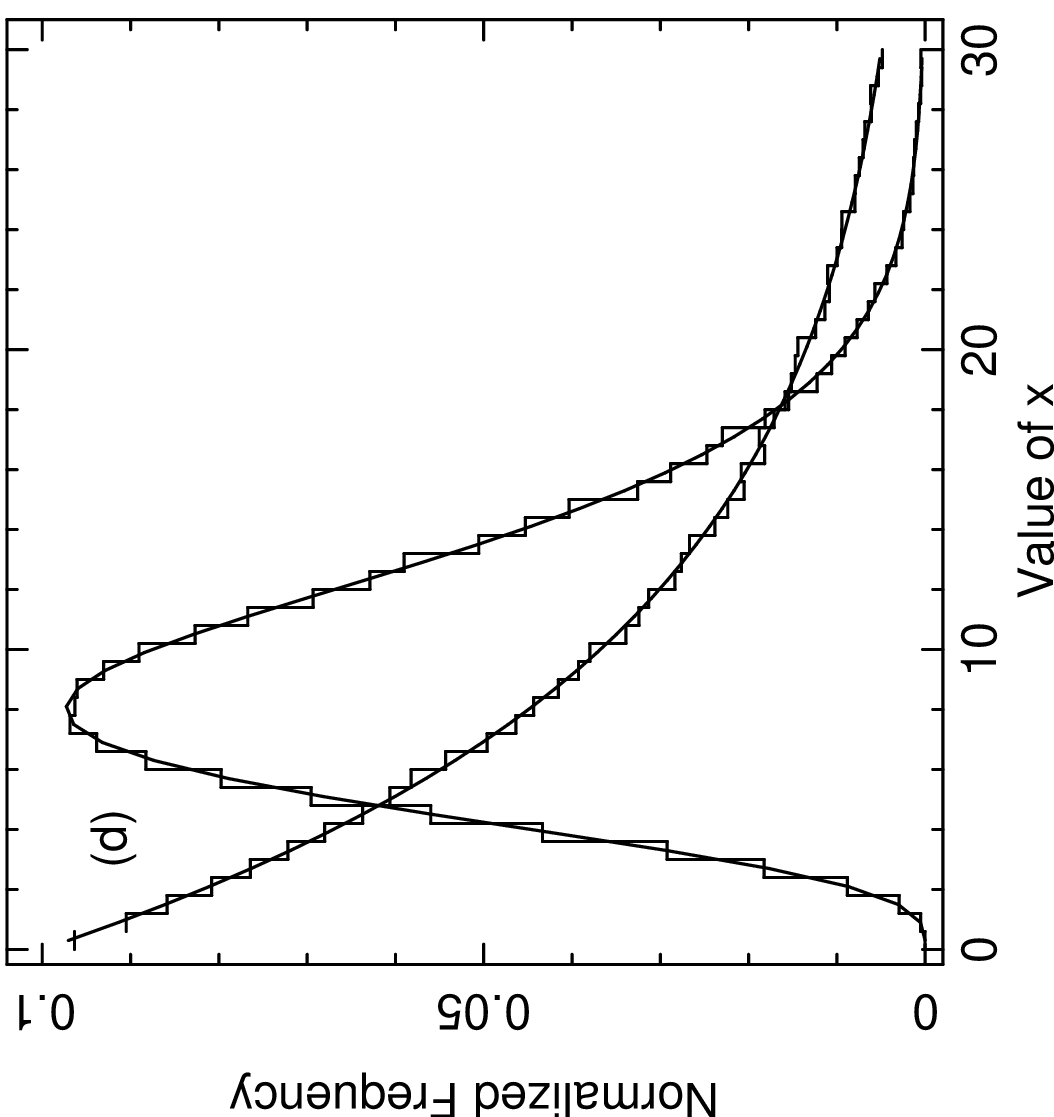}
\end{center}
  \caption{\footnotesize Illustration of the relationship between standard normal, $\chi^2$ and exponential variables. Using the event list of the time series presented in Figure\,\ref{f:transient}, and calculating the phases corresponding to the independent periods between 0.5 and 3600\,s, in panel (a) we see the normalized frequency distribution of the scaled variable $c=\sqrt{2N}C$ overlaid with the standard normal density function; in panel (b) we see its square, $c^2=2NC$, overlaid with the $\chi^2_1$ density function; and in panel (c) we see the normalized distribution of the Rayleigh statistic, $R^2=c^2+s^2$, overlaid with the $\chi^2_2$, which is the exponential density function with $\tau=2$. Panel (d) illustrates using $10^5$ pseudo-random numbers, the difference between the sum of five $\chi^2_2$ variables, which results in the unimodal a $\chi^2_{10}$ distribution peaking at eight, and the distribution that results from multiplying or scaling a $\chi^2_2$ by five, which yields the exponential distribution with $\tau=10$.
\label{f:variables}}
\end{figure*}

Here as in the other cases, every single additional measurement brings an additional value of $n_i$ in channel $i$ from which the accuracy of estimates of $\nu_i$ can be improved. However, unlike the procedure for independent pixels or spectral channels, the joint likelihood function is used to determine the most likely model parameter values given the available data, and this is done by fitting using the $C$ statistic \citep{1979ApJ...228..939C} given by $C=-2\ln L$, with $\ln L$ given by Equation (\ref{eq:poissonLogLikelihood-model}).
Since fitting entails varying the values of the parameters in order to maximize the likelihood and thus minimize the value of the fit statistic, comparing different model parameters is done through the ratio of likelihoods that translate to differences of log-likelihoods. Hence, terms (in this case \emph{the} term) that does not depend on the parameters, will not contribute and can be dropped. The fit statistic becomes
\begin{equation}
C = 2\sum_i \left[ \nu_i - n_i\ln({\nu_i}) \right].
\end{equation}
Thus what is updated with each measurement or iteration are the model's parameter values, and the identification of transient spectral features is based on the likelihood of the current freshly updated observed spectrum in relation to the previous best fit model. The thresholds are optimized for the application, both in terms of the individual value of the likelihood ratio after each measurement and in terms of the number of consecutive unlikely values. A discussion of the historical context, the motivation for, and the details that relate to procedures used in comparing data and models is the subject of another paper in preparation, and we therefore do not go any further into this here.

\subsection{Transients in Power Spectra}
\label{s:powerSpectra}

The power spectrum is estimated by the periodogram and carries information about various scales in the observed system. Because it presents what is called ``power'', which is a measure of the amount of ``activity'' at a given frequency or timescale, the power spectrum conveys information about both dynamical and distance scales. Any spontaneous and stochastic, or triggered and stimulated event that leads to a reconfiguration of the dynamics of the system, be it local or global, will generally manifest itself in the power spectrum. How this will then translate into the periodogram for a given geometry and observing conditions depends upon how large, long lasting, and coherent the event, and surely most importantly, on the sensitivity of the instrument with which the system is being observed. Nevertheless, such transients can potentially be detected and identified as such, and we seek the best means to do so.

The approach most resembles the treatment of energy spectra in its details. For power spectra as well, the problem can be treated as though the values in each channel, in this case frequency channels, were independent of one another---something that is only true when the power spectrum is globally flat with equal power at all frequencies---or it can be treated with the use of a global model that prescribes how the values in the different channels are related. Note that in both cases, the model can be theoretical, or empirical and constructed from the successive iterations of measurements and refinement of the estimated spectrum by computations of the periodogram. Therefore, conceptually and procedurally, this problem is the same as for the energy spectrum. 

There are two differences, however. One is superficial: that models for power spectra are generally fewer, largely phenomenological and often simpler than those used for energy spectra. The other is fundamental: that the values of power estimates in frequency channels are distributed neither as Poisson nor as normal variables, but are instead related to $\chi^2$ and exponential distributions. The reason is that each power estimate is calculated from a sum of squared standard normal variates. 

For an ensemble of detected events, each with its own arrival time, the simplest way to calculate the power of the fundamental harmonic at a given frequency $f$, is to map each arrival time $t_i$ to its phase $\phi_i$ within the periodic cycle that corresponds to that frequency ($p=1/f$), and calculate the Rayleigh statistic \citep{1983ApJ...272..256L}:
\begin{equation}
	R^2 = 2N(C^2 + S^2)
\end{equation}
where $C$ and $S$ are defined as:
\begin{equation}
\label{eq:CandS}
	C = \frac{1}{N} \sum_{i=1}^N \cos{\phi_i}
	\hspace{4mm} {\rm and} \hspace{4mm}
	S = \frac{1}{N} \sum_{i=1}^N \sin{\phi_i}. 
\end{equation}

First, the expectation value of the functions $\cos\phi$ and $\sin\phi$ is zero: $\mean{\cos{\phi}} = \mean{\sin{\phi}} = 0$. Therefore, so are those of $C$ and $S$.  Second, the variances of $\cos\phi$ and $\sin\phi$ both equal one half: ${\rm V}[\cos{\phi}]={\rm V}[\sin{\phi}]=1/2$. Therefore, those of $C$ and $S$ are a factor of $N$ times smaller: ${\rm V}[C]={\rm V}[S]=1/2N$.  Finally, since ${\rm V}[mX]$\,=\,$m^2 {\rm V}[X]$, where $m$ is a numerical constant, the scaled variables $c$\,=\,$\sqrt{2N}\cdot\/C$ and $s$\,=\,$\sqrt{2N}\cdot\/S$ have a variance of one: ${\rm V}[\!\sqrt{2N}\cdot C] = {\rm V}[\!\sqrt{2N}\cdot S] = 2N\cdot {\rm V}[C] = 1$. 

Note however, that the phases are uniformly distributed between 0 and 2$\pi$, and the sine and cosine are distributed between $-1$ and 1 with their characteristic, symmetric U-shaped distribution with minimum at 0, and rising slowly toward the edges where it peaks very sharply. It is the summing and averaging of several identically distributed values that yields the two normal variables $C$ and $S$, and standard normal $c$ and $s$.

\begin{figure*}[htb]
\epsscale{1.0}
\begin{center}
\includegraphics[scale=0.3, angle=-90]{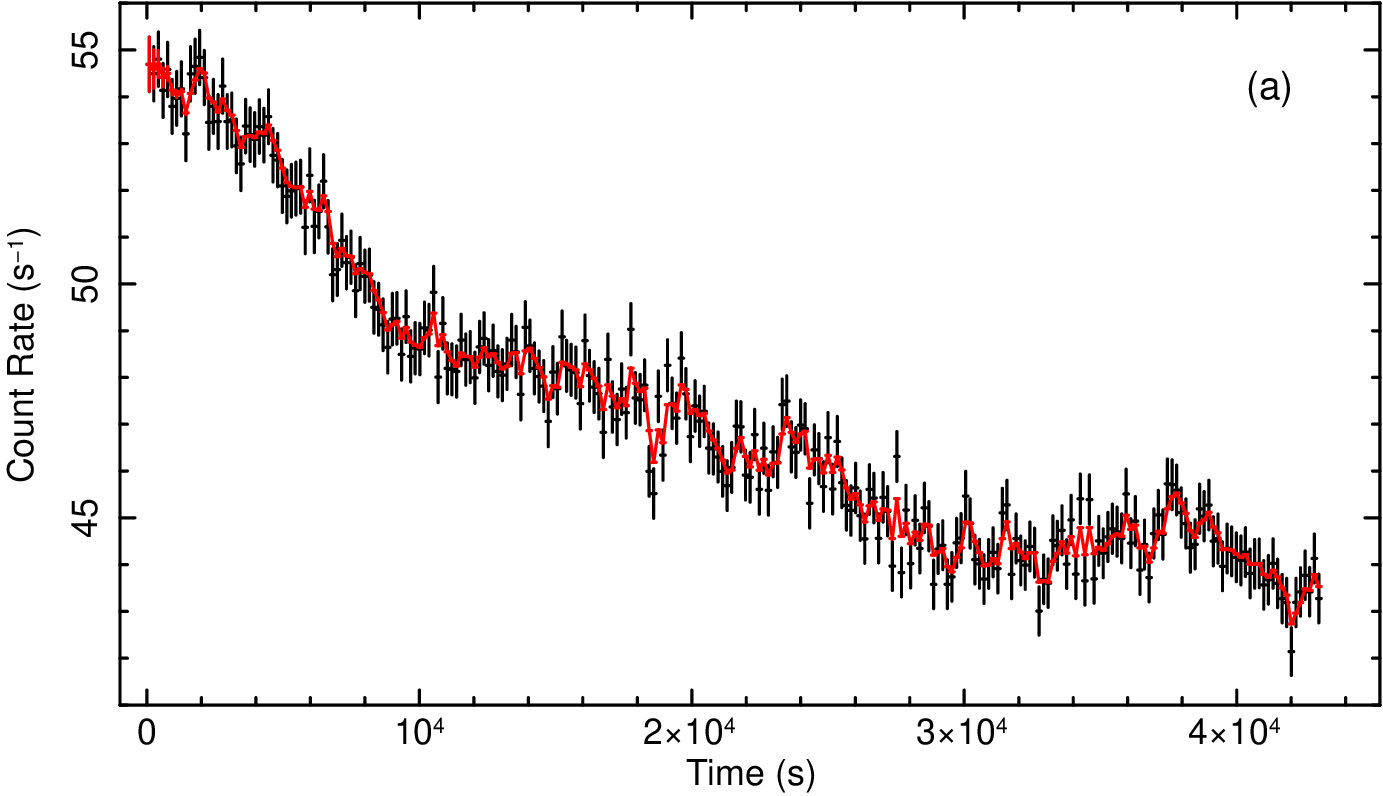} \hspace{2mm}
\includegraphics[scale=0.225, angle=-90]{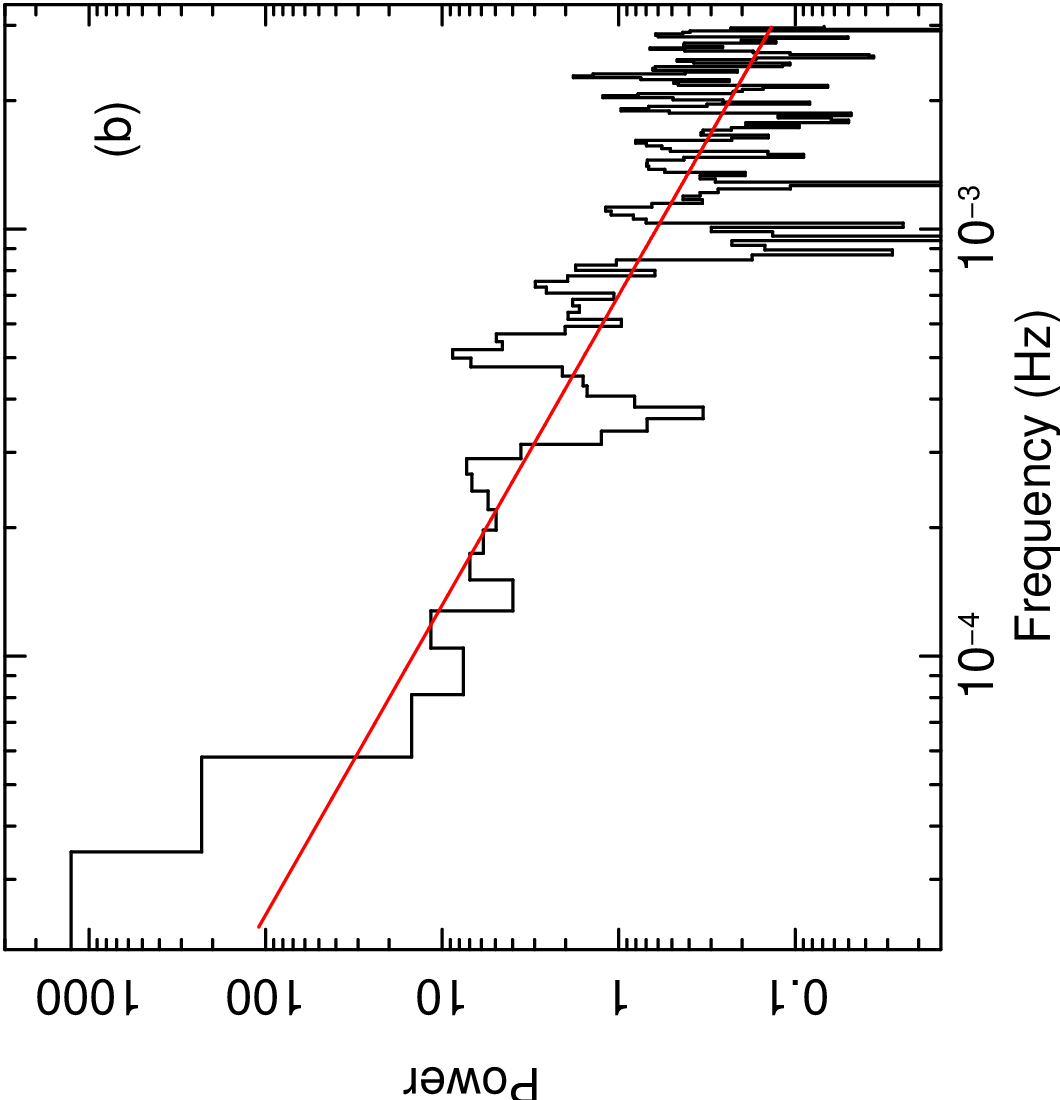} \hspace{2mm}
\includegraphics[scale=0.225, angle=-90]{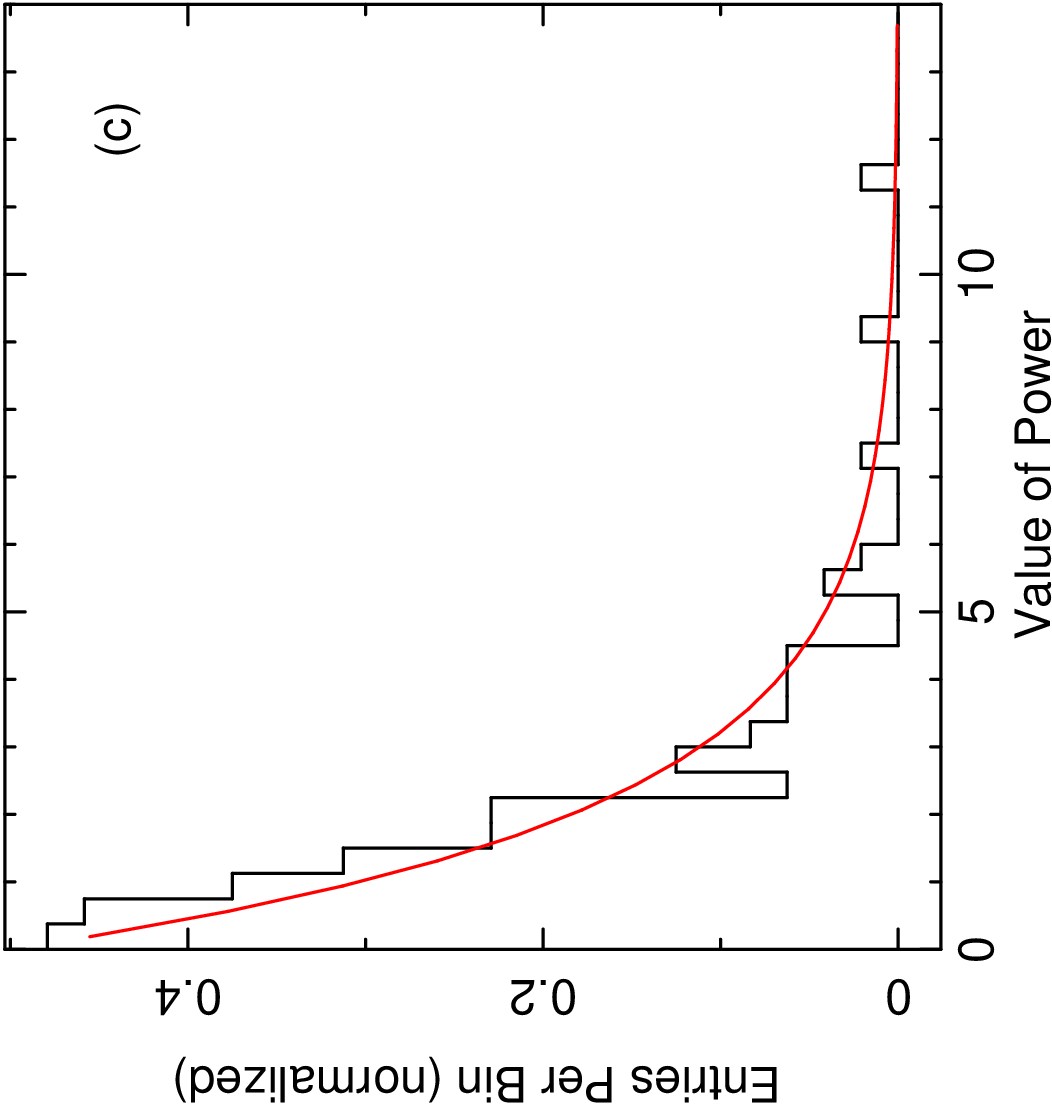}
\end{center}
  \caption{\footnotesize Illustration of periodogram powers of astrophysical red noise as scaled $\chi^2_2$ variables using a 43 ks observation of Mkn 421 \emph{XMM-Newton}. Panel (a) shows the Reflecting Grating Spectrometer time series in rates (0.3--2 keV with 85\,s bins). In black are the measured data and in red are those that result from applying a Kalman filter \citep{Kalman:1960kh}, which very effectively suppresses the white noise scatter \citep[see also][]{1997A&AS..124..589K}, and thus allows for a better constrained fit on the resultant periodogram that is shown in panel (b) with the best fit power-law model. Panel (c) is the distribution of de-trended periodogram powers overlaid with the analytical form of the $\chi^2_2$ density function. The excess power at the lowest frequency, about 12 times above the best-fit in panel (b), is due to the global trend in the time series marked by a large difference between the start and end of the observation, and is seen in the right hand tail and greatest value in the histogram in panel (c). (A color version of this figure is available in the online journal.)
\label{f:rednoisepowers}}
\end{figure*}

This implies that: $$R^2=c^2 + s^2=2NC^2 + 2NS^2=2N(C^2 + S^2)$$ is the sum of the squares of two standard normal variables. Squaring a standard normal yields a $\chi^2$ variable with one degree of freedom (dof). Summing $\chi^2$ variables yields another $\chi^2$ variable with a number of dof that is the sum of the dof of the variables being summed (illustration in \cref{f:variables}). 
Therefore, the power being the sum of two $\chi^2_1$ variables is $\chi^2_2$ distributed with a mean and standard deviation of two (variance of four). This is convenient due to the simplicity of the purely exponential $\chi^2_2$ density function:
\begin{equation}
\chi^2_2(x) = \frac{1}{2} e^{-x/2}.
\end{equation}

The caveat here is that this is only true if the power estimates at different frequencies are independent, which is only true for non-variable processes: the same kind that lead to a globally flat power spectrum with equal power at all frequencies. This is therefore an ideal case that should be treated as such, and not given more attention than it deserves. Nonetheless, it is instructive and illustrates how normal, $\chi^2$ and exponential probability densities can be related.

A natural choice for a general $\chi^2$ fit statistic is twice the negative of the log-likelihood of Equation (\ref{eq:chiSquareLogLikelihood}), $K=-2\ln L$,  and dropping the term that does not depend on the parameters $k_i$ yields:
\begin{equation}
K = -2\sum_i \left[ \left(\frac{k_i}{2}-1\right)\ln x_i - \frac{k_i}{2}\ln 2 - \ln\Gamma\left(\frac{k_i}{2}\right) \right].
\end{equation}

Just as the $C$ statistic is optimal for Poisson distributed data because it is derived from the likelihood function, which is in turn derived from the probability density for Poisson variables, the $K$ statistic is optimal for $\chi^2$ distributed data for the same reason. It is optimal for fitting a model to a set of measurements that are samples of  random $\chi^2$ variables with potentially different dof $k_i$ in each channel $i$.

Now, a much more common and also more general case than that of the flat spectrum of a non-variable process, is that of a power-law distribution of power estimates as a function of frequency usually called red noise in Astrophysics (see the highly sophisticated work by \cite{2010MNRAS.402..307V} on a Bayesian approach to evaluating the significance of an almost or strictly periodic signal in red noise for a discussion of its properties). The ideal case of a globally flat power spectrum considered above is the simplest manifestation of a red noise process: that with a spectral index of zero. The power values in red noise are related to one another in a well-defined manner through the relation Power $\propto f^{-\alpha}$, where $f$ is the frequency and $\alpha$ is the power spectral index. This is a model that describes the shape of the underlying power spectrum; not the shape of any particular periodogram that serves as an estimate of it, subject to various degradation factors, distortions and statistical fluctuations. 

As is the case for the likelihood, the absolute value of the power in the periodogram is not meaningful: it is only the relative power at a frequency compared to that at another that carries meaning. Therefore, any normalization factor appropriate for the application can be used to scale the periodogram up or down. The key, however, is that we are working with \emph{scaled $\chi^2_2$ variables}.  This is demonstrably true for astrophysical red noise (\cref{f:rednoisepowers}), because dividing the power estimates in each channel by the best-fit power-law model yields values that are $\chi^2_2$ distributed.\footnote{It is necessary to demonstrate this relationship using real data, because many algorithms used to generate power-law noise, as the one commonly used by \cite{1995A&A...300..707T}, are precisely like this by construction, for the Fourier components are generated by multiplying the model spectrum at each frequency by pseudo-random standard normal variates, one for the phase and one for the amplitude. Squaring and summing to get the power and then dividing by the best-fit power-law model will always give $\chi^2_2$ distributed powers.}
This would also be true for \emph{any} power spectral shape \emph{if} the process can be considered as one of simply scaling the basic $\chi^2_2$ variable that results from summing two squared standard normal variables by the underlying model, whatever the shape. This is indeed what we have always seen in our work in the frequency domain, and thus assume this to be generally the case.\footnote{\cite{1986ssds.proc..105D} and \cite{1990ApJ...364..699A} discuss this issue and also adopt this position.}

\begin{figure*}[htb]
\epsscale{1.0}
\begin{center}
\includegraphics[scale=0.3, angle=-90]{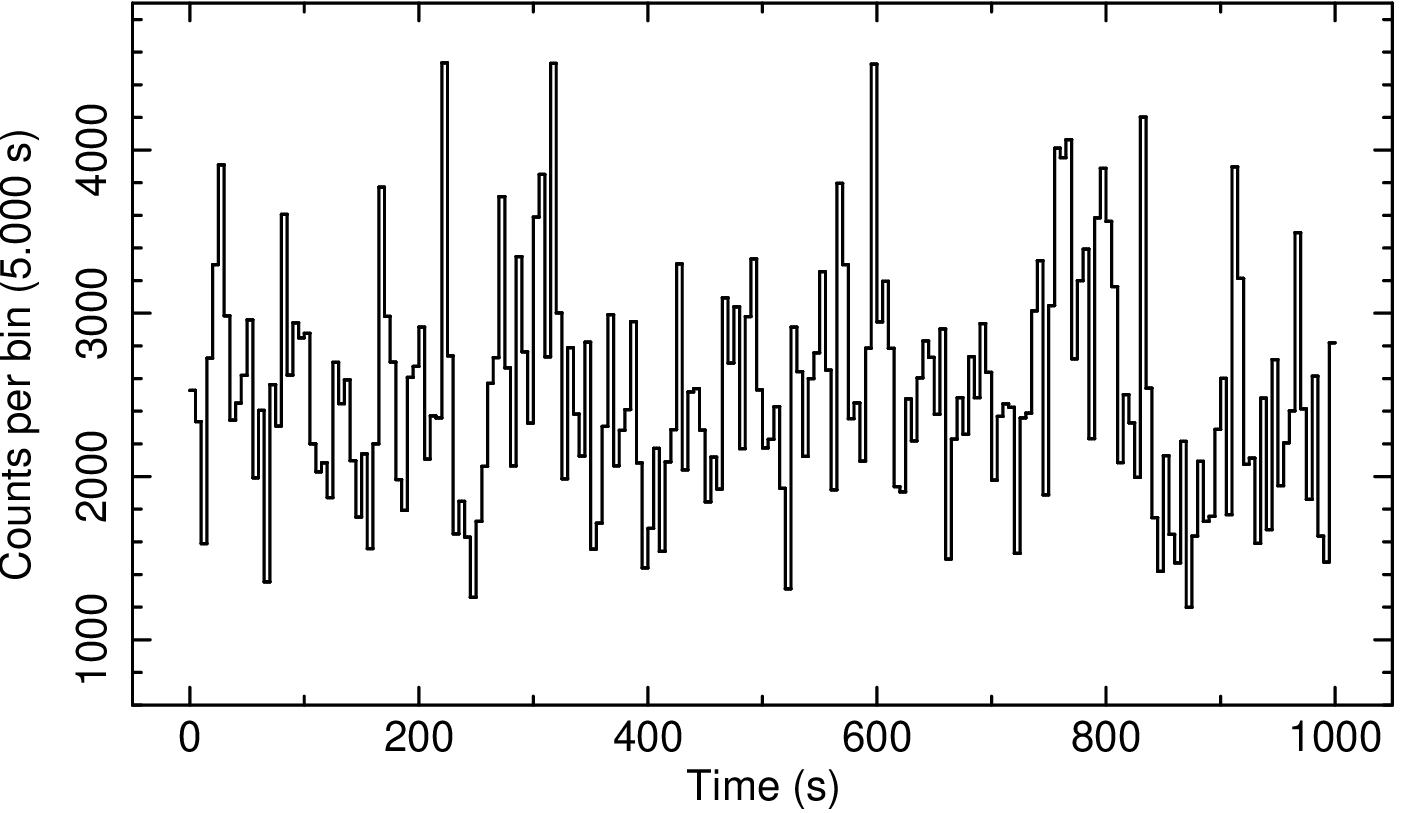} \hspace{1mm}
\includegraphics[scale=0.225, angle=-90]{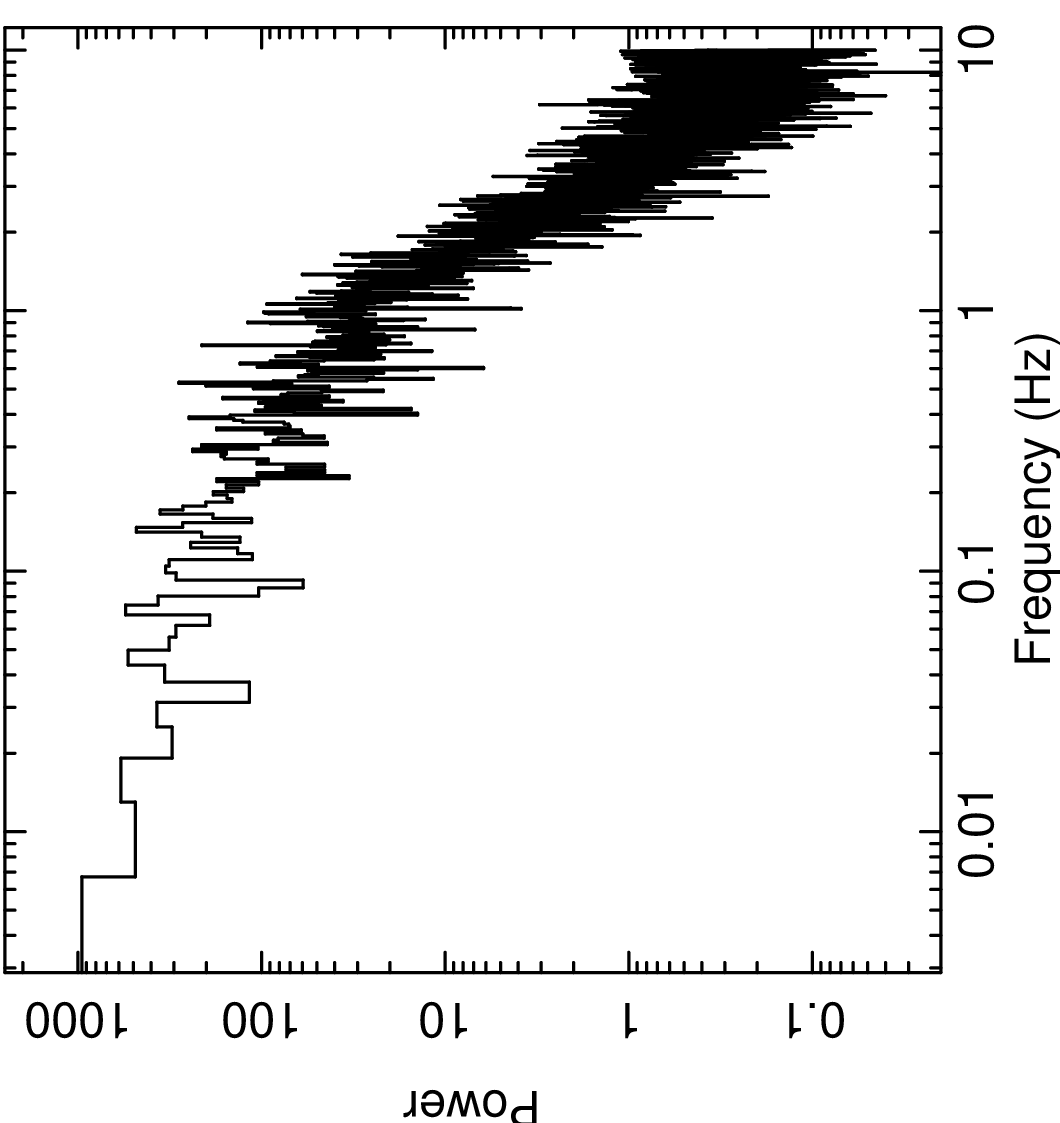} \hspace{1mm}
\includegraphics[scale=0.225, angle=-90]{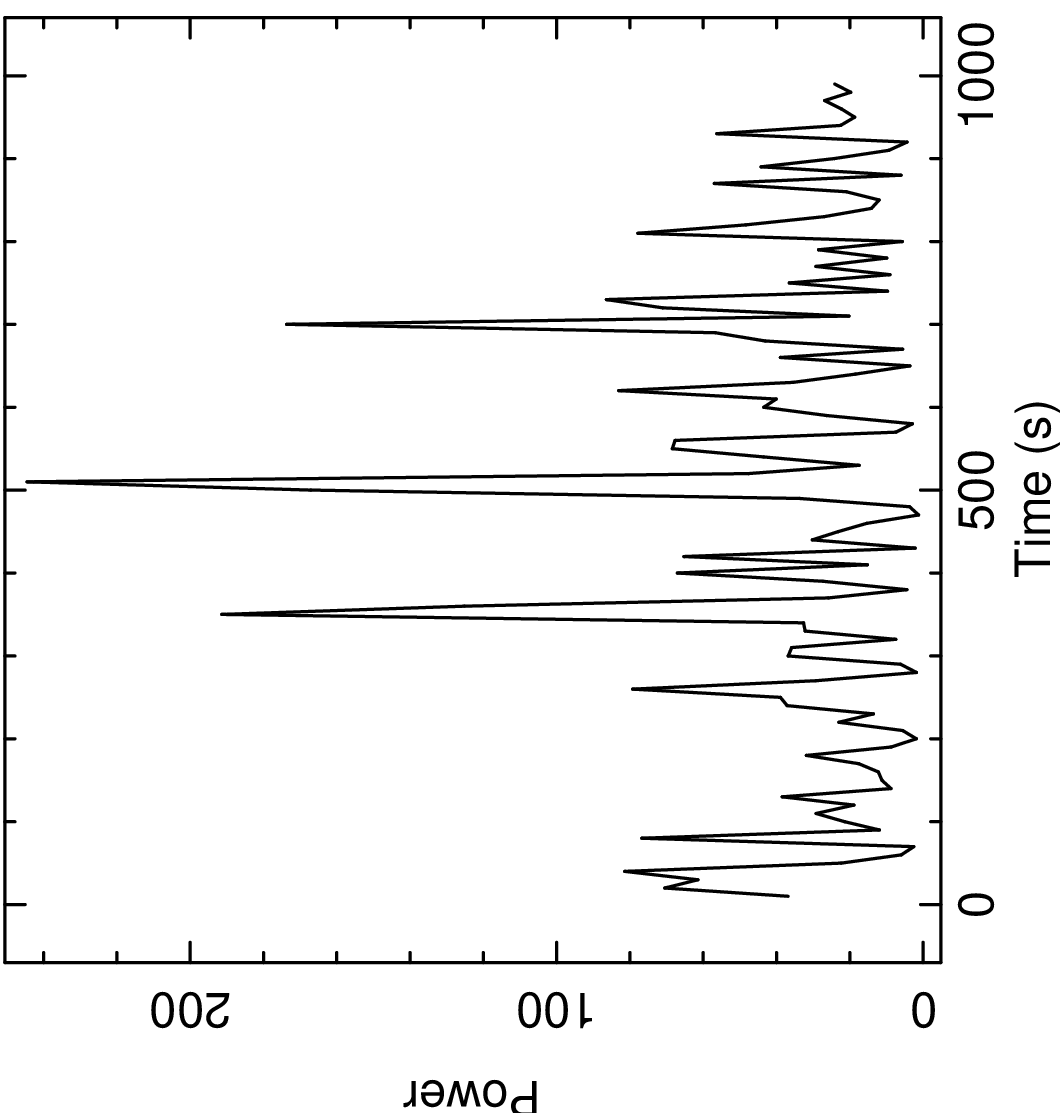} \\\vspace{1mm}
\includegraphics[scale=0.3, angle=-90]{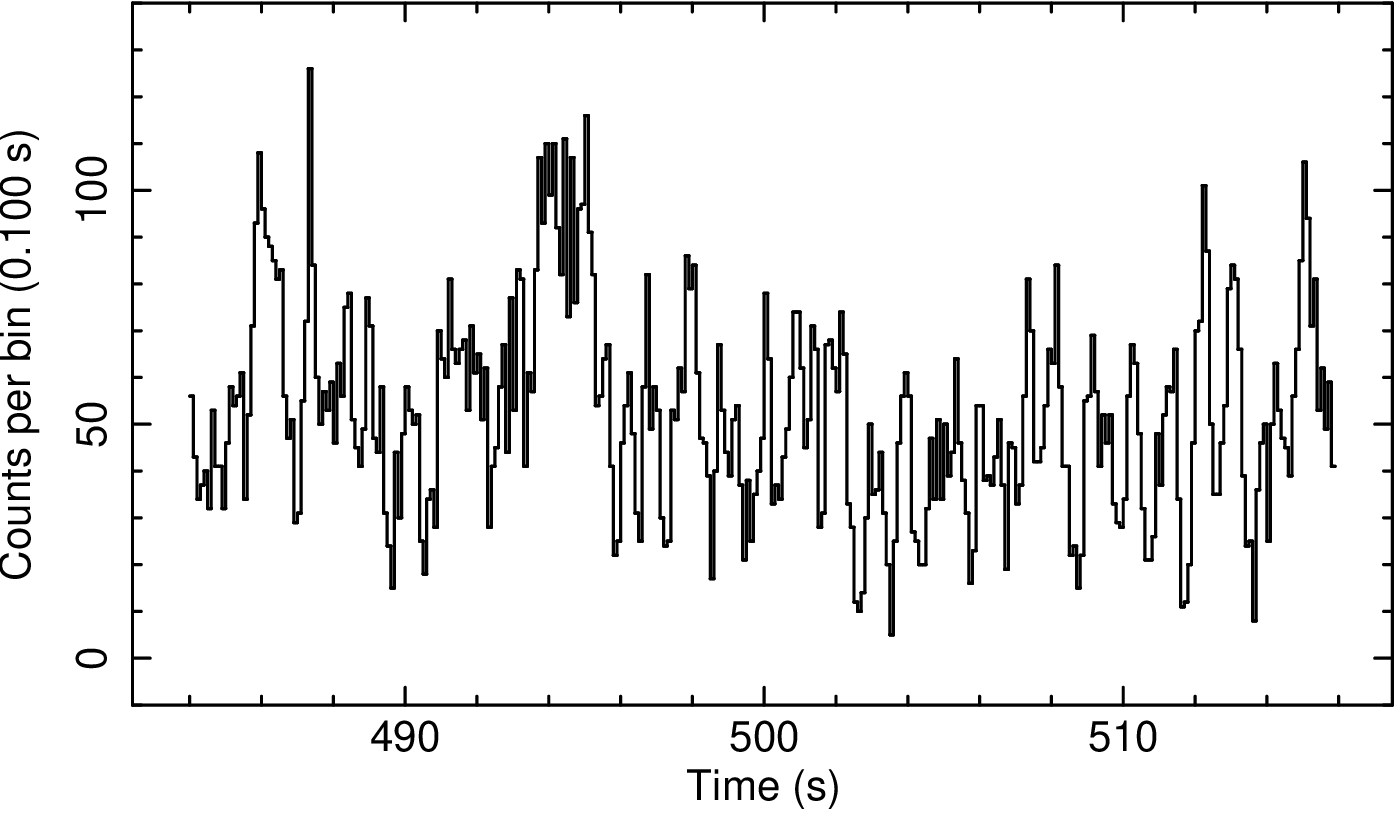} \hspace{2mm}
\includegraphics[scale=0.225, angle=-90]{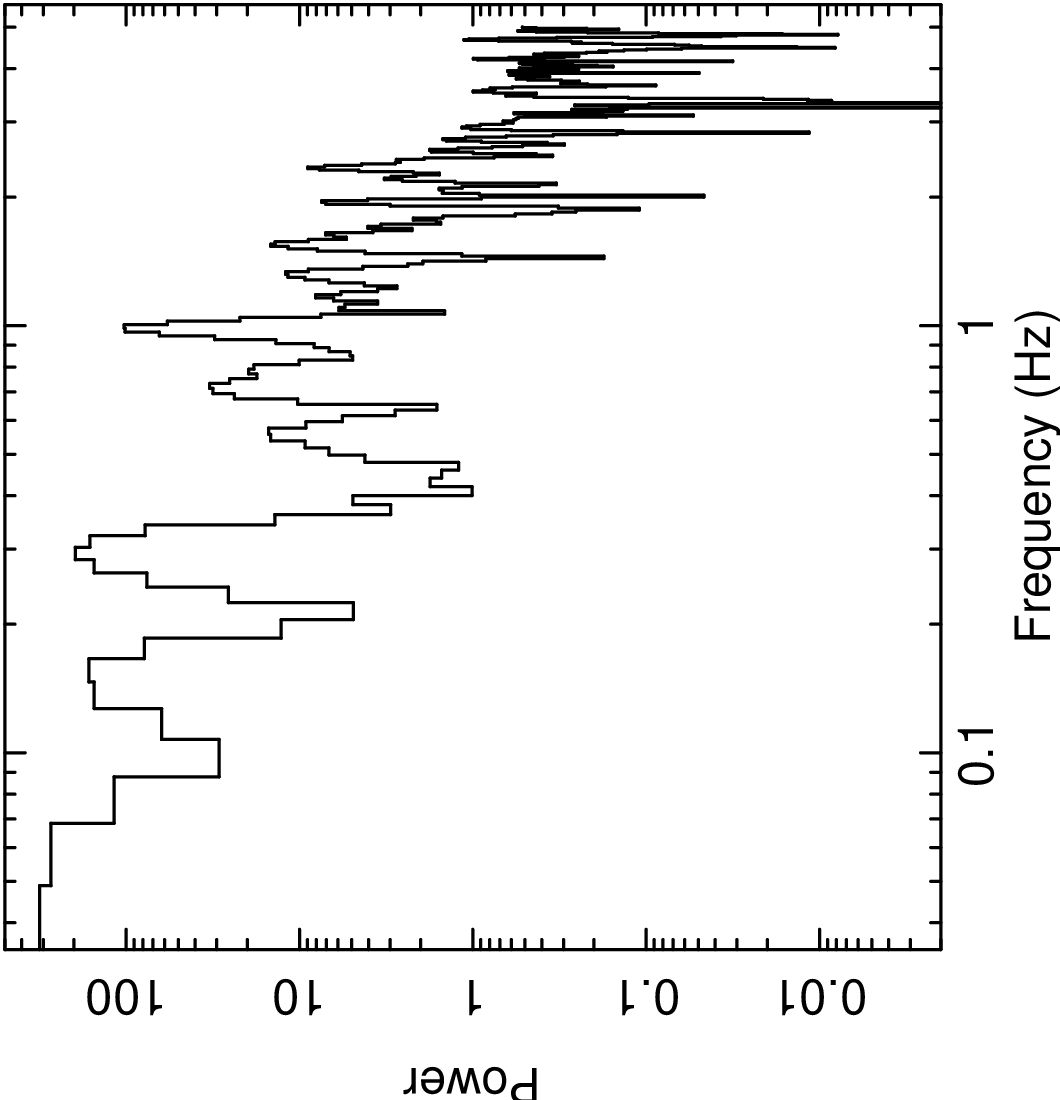} \hspace{2mm}
\includegraphics[scale=0.225, angle=-90]{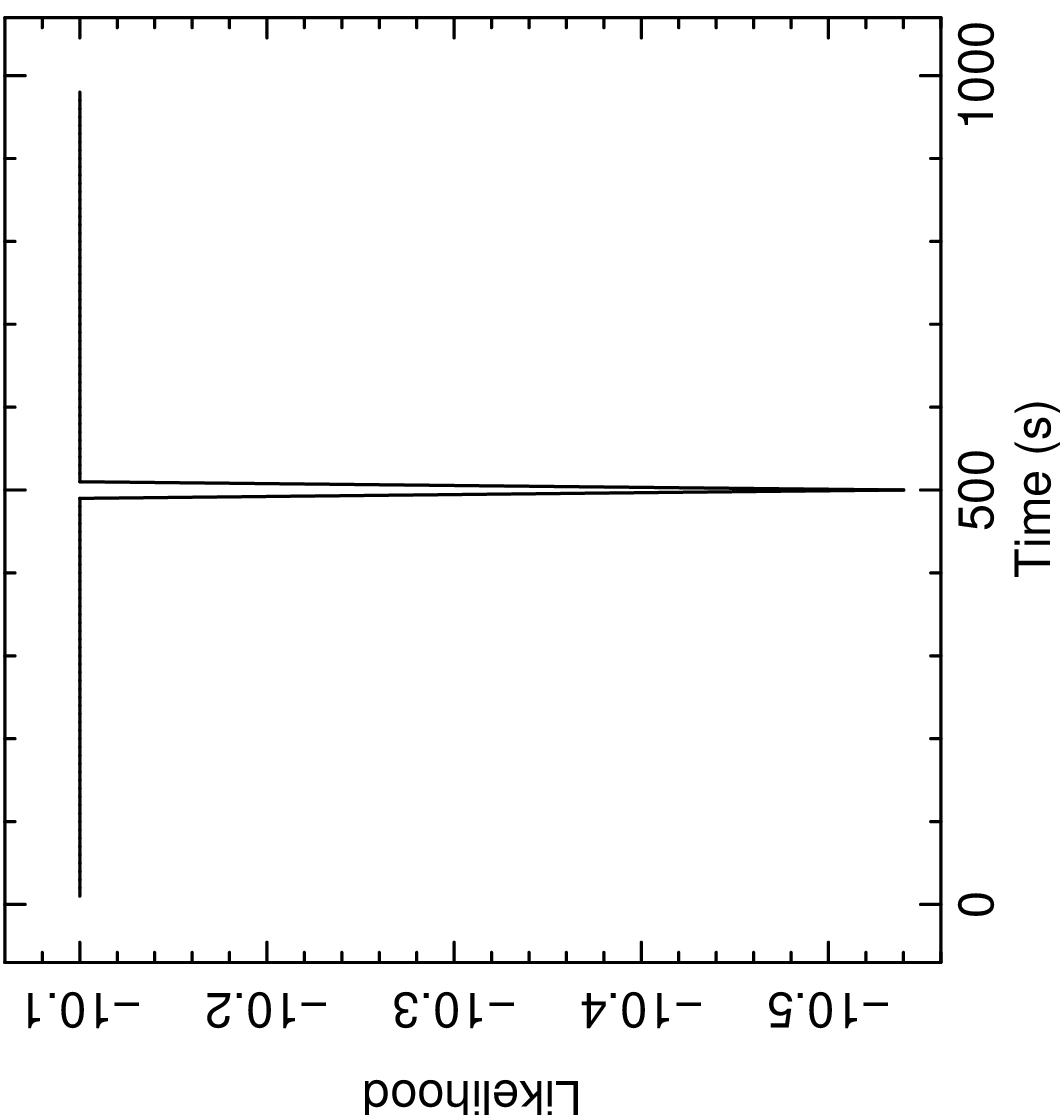}
\end{center}
  \caption{\footnotesize The top row shows the time series of the entire observation (binned to a resolution of 5\,s for clarity of presentation); the periodogram made from the Kalman-filtered, 0.05\,s resolution time series of the event arrival times; and the power at 1 Hz estimated at 10\,s intervals from the Rayleigh periodogram of the event arrival times as a function of time. The bottom row shows a zoom on the time series during the transient QPO from its start at 485\,s until its end at 515\,s after the start of the observation; the periodogram of the Kalman-filtered 0.05\,s resolution time series; and the log-likelihood as a function of time where only detections beyond the established threshold are shown. The QPO is characterized by 30 cycles of an almost periodic signal centered on 1 Hz with a standard deviation of 1/20 about that frequency and a pulsed fraction of 27\%.
\label{f:transientInPowspec}}
\end{figure*}

Therefore, whether or not we have a model describing the power spectral shape, we work with the power estimates at a given frequency as though they were $\chi^2_2$ variables scaled differently in each channel. This implies they are distributed according to the exponential density function (\ref{eq:exponentialDensity}), that their joint likelihood function is given by Equation (\ref{eq:exponentialLikelihood-model}), and thus the log-likelihood by Equation (\ref{eq:exponentialLogLikelihood-model}): $$\ln L(\pmb{\tau} | \pmb{x}) = -\sum_i (\ln\tau_i + x_i/\tau_i).$$
In the periodogram, $x_i$ is the measured, and $\tau_i$ is the model-predicted power in frequency channel $i$.  We can therefore construct another fit statistic specifically for fitting periodograms based on this expression  (\cite{1986ssds.proc..105D} also derive and use this statistic; see also \cite{2012ApJ...746..131B}) as was done above with $K$ for the general case of different $\chi^2$ variables, but now for the general case of a collection of exponentially distributed variables, such that $B = -2\ln L$:
\begin{equation}
\label{eq:exponentialLogLikelihood-global}
B =  2\sum_i (\ln\tau_i + x_i/\tau_i).
\end{equation}
When working with a power spectral model, the $B$ statistic is used to continuously compare the observed with the predicted distribution of powers in the periodogram, as is done with the $C$ statistic in the case of Poisson distributed events in an energy spectrum, or the $K$ statistic for $\chi^2$ distributed measurements more generally. Sensitivity to detect changes in the overall shape of the spectrum increases quickly with the number of iterations. However, fluctuations in the power estimates in each channel always remains important due to the intrinsically high variance of exponential variables for which the standard deviation is equal to the decay constant ($\mu=\tau,$ $\sigma^2=\tau^2$ and thus $\sigma=\tau$).

As a demonstration, we consider a hypothetical observation in X-rays of a bright (500\cps) accreting system whose variable emission comes mostly from two components: the accretion disk, and the hot and turbulent gas in the inner flow. In both, the emission processes are connected on all timescales, and thus each gives rise to a red noise component. The accretion disk is much larger in extent and has a sharp inner radius. It dominates at lower frequencies with a power-law index $\alpha=-1$, and has a high-frequency cutoff beyond which it does not contribute to the power spectrum. The turbulent inner flow is much smaller in extent because it is bounded by the inner edge of the disk. Its emission is more variable and dominates the high-frequency part of the spectrum with a power-law index $\alpha=-3$. 

In this case, we are interested in monitoring the range of frequencies between 0.1 and 10 Hz for the appearance of a weak, short-lived, transient QPO that we expect to appear at or very near the break in the power spectrum at 1 Hz, which marks the boundary between the disk and the turbulent inner flow. For this, we make a periodogram every 10\,s with the events accumulated during this time interval, and monitor the power at one or any number of frequencies. Because we are interested in a short-lived QPO, the strategy is different than for the previous example of the time series: we cannot rely on the transient persisting in more than one ``measurement'', and therefore must establish a single detection threshold that is constraining enough for our application. This threshold is established using simulations. We have done this for the power at 1 Hz, the frequency of interest for us here, to first determine the average expected power (35), and then establish a threshold (log-likelihood of $-10.1$, and thus a likelihood of $4.1\times 10^{-5}$) that ensures a level of false detections of 5\%.

The observation and the analysis are presented in Figure \ref{f:transientInPowspec} in which we see  that the transient QPO is clearly detected in the likelihood monitoring, but, as should be expected from its very short lifetime, is not at all evident in the periodogram of the entire observations, even though it is indeed present. It is important to emphasize once more that to maximize the power of this method, the detection strategy must be devised and optimized for the application.

\section{Concluding Remarks}
\label{s:conclusion}

Treating and interpreting data as statistical evidence in seeking to further our understanding of physical processes and of the behavior of complex systems such as those we observe in astrophysics, using all the information carried by these data, is most directly done through the use of the likelihood function appropriately chosen according to the statistical distribution of the random variable that is measured. This approach yields a most powerful and effective means for treating any kind of data, because no information about the process under investigation is left out or ignored, nor degraded in any way; everything is seamlessly taken into consideration and into account in the statistical decision-making process. This is particularly appropriate for the search and identification of transient phenomena in all data domains of interest in astrophysics (temporal, spatial, energy, and frequency), and this is what we have tried to demonstrate.

The method presented is well suited to handle the first two classes of transients that have a non-variable background, be it nil or of some constant level, without any further considerations. Evidently, in this as in any other method, the identification efficiency ultimately depends very intimately on the relative strength of the transient signal. Identifying transients with the described procedures is perfectly suited for analyzing archival data, where the data sets are already complete and well defined. It is, however, also powerful for real-time applications, maybe especially so. Handling the third class of transients characterized by a variable background requires additional care, as discussed in the Section\;\ref{s:identifyingTransients}. Here are some additional considerations:

If the process is variable but predictable as with a sinusoidal modulation, for example, then this is a simple extension of the procedure using a model, but in which it itself evolves as a function of time; the formalism is otherwise identical. If the process is variable and unpredictable, it implies that the measurements in each pixel or channel are not distributed according to a particular and unchanging probability distribution. Instead, even though it may belong to a given family of distributions based on the nature of the process, it will inevitably have a hybrid shape due to the changes in that process as a function of time which we must model empirically. Therefore, each pixel or channel is treated individually, but because we have no a priori expression for the likelihood function, the intensity and how it is distributed can be characterized approximately using the running mean and variance. 

For highly variable processes, where deviations in shape from known probability distributions are large, looking at the distribution of measured values per pixels or channel does not make much sense because the changing intensity in each is like a very small window onto complex interactions that give rise to variable emission that cannot be described by random variables with stationary probability distributions. Doing this is very limited in usefulness even when it is possible to find a function that can be fitted to the distribution of measured intensities such as a log-normal distribution, for example. However, a variable process can be highly non-stationary in the time domain, but stationary in frequency space, having a power spectrum that does change in time. This is analogous to a variable point source whose intensity varies markedly in successive images but whose location in the sky remains the same, and whose shape as it appears in each image is as always given by the instrument's PSF. Combining the information carried by the data in the time and frequency domains, and treating it simultaneously in the fashion described in this paper is certainly a most powerful means of analysis for detecting transient features in highly variable processes. 

Note that, as alluded to in Section\;\ref{s:identifyingTransients}, the crucial elements in working with variable processes are the timescales involved: in this application, that of the transient with respect to that of the variability. More specifically, since the stationarity of the probability distribution can be considered as being a function of the timescale at which the process is viewed, in general it is possible to have a running estimation of that probability distribution which is stationary up to a given timescale, but evolves on longer timescales. In this way, the likelihood function and all the associated statistics are well defined at any point in time, and the method becomes a more general, time-dependent form of the procedure presented. As stated, details relating to this will be presented elsewhere.

The generality and wide applicability of the method we presented in this paper, can also be viewed as its primary limitation. This is not unlike the majority of statistical methods with a specific purpose, and is related, very simply in this case, to the fact that there are various kinds of transients, and we want to detect them all with maximum sensitivity. Therefore, as was shown in both the case of a transient in a time series and in a power spectrum, the power of the method relies on simulations for an accurate estimation of the statistics of the process, and for defining the detection thresholds, that will in general be geared toward a certain class of events. Nonetheless, there are in principle no limits to the number of automated searches that can be performed in parallel, for any given data set or application. Furthermore, the generality of the formalism is such that it can be applied to identifying transients in other parameter spaces, where the independent variable is not time.

Another limitation relates to the full reliance on well-defined statistics because the likelihood function otherwise simply cannot be constructed. Although this may not appear to be an important factor in many methods commonly used, the truth is that it always is, but it is not necessarily recognized because of the generalized implicit assumption of normal statistics. The periodogram is an excellent example of the importance of using the correct probability distribution. 

Having recognized that the power values in a frequency channel of any periodogram are exponentially distributed with a mean given by the expected power in that channel, the one-sided tail probability of finding a power value of 60 or greater, for instance, when the expectation is 30, is 0.135 or 13.5\%, which everyone would agree is definitely not worthy of even a second glance in terms of ``statistical significance''. However, using normal statistics (mean power of 30 and standard deviation of $\sqrt{30}$, say), finding a value of 60 or greater is a 5.47$\sigma$ result with a probability of about $10^{-8}$.

Therefore, although this limitation could be a weakness of the method in certain circumstances, it is clearly also a strength that highlights a very fundamental point which should rightly be incorporated in any statistical method: that of using the correct statistics. We believe this likelihood-based approach is a way forward, and will prove to be very valuable in many applications where transients and outliers are of importance.

\acknowledgements{The author is grateful to Andy Pollock for many stimulating and inspiring discussions, to Marnix Bindels for his suggestions and help in Java programming, especially with the AggregatePeriodogram, and to Joel Velasco for his reading recommendations in the philosophy of statistics and especially \emph{Statistical Evidence} \citep{Royall:1997vc}. Tom Maccarone, Michiel van der Klis, and Jeremy Heyl made comments and posed questions that helped improve the manuscript.}

\appendix

\section{Probability Density and Likelihood Functions of Poisson, normal, $\chi^2$, exponential and inverse-exponential Variables}
\label{a:equations}

\subsection{Poisson Variables}
The Poisson probability density function is given by
\begin{equation}
\label{eq:poissonDensity}
f(n;\nu) = \frac{\nu^n e^{-\nu}}{n!},
\end{equation}
where $n$ is a random variable representing the number of detected counts during a specific time interval---a positive integer; and $\nu$ is the rate parameter that represents the number of counts expected in the same time interval---a positive real number. The likelihood function for a single measurement is given by the density function, as proportional to it in the sense described in Section\;\ref{s:likelihoodFunction}:
\begin{equation}
\label{eq:poissonLikelihood-single}
L(\nu|n) \propto f(n;\nu) = \frac{\nu^n e^{-\nu}}{n!}.
\end{equation}

The semicolon is used to separate $n$ and $\nu$ to indicate that $\nu$ is a parameter and not a variable, and that its value must be fixed in order to know the form of a specific density function. It is not a conditional probability of a bivariate normal density function, for example, where we fix the value of $y$ and look at the density function for $x$, and is written $f(x|y)$. To write the likelihood function, we use $L(\nu|n)$ to indicate that it is a function of the parameter $\nu$ given the measured data that is now fixed. We generally use ``$=$'' instead of ``$\propto$'' for simplicity.

For more than one measurement, a vector of measurements $\pmb{n}$ where each is now identified by the subscript $i$, the likelihood function is
\begin{equation}
\label{eq:poissonLikelihood-joint}
L(\nu|\pmb{n}) = \prod_i \frac{\nu^{n_i} e^{-\nu}}{n_i!}.
\end{equation}
If we are instead working with a model and comparing the measured number of events with what the model predicts in each channel of a spectrum, for example, then each measurement $n_i$ is compared with the model-predicted parameter $\nu_i$, and the expression for the joint likelihood becomes
\begin{equation}
\label{eq:poissonLikelihood-model}
L(\pmb{\nu}|\pmb{n}) = \prod_i \frac{{\nu_i}^{n_i} e^{-\nu_i}}{n_i!}.
\end{equation}
It is often convenient to work with the natural logarithm of the likelihood function in which products become sums, ratios become differences, and powers become products. The log-likelihood function for Poisson variables is therefore given by:
\begin{equation}
\ln L(\pmb{\nu}|\pmb{n}) = \sum_i \left[ n_i\ln({\nu_i}) - \nu_i - \ln(n_i!)\right].
\label{eq:poissonLogLikelihood-model}
\end{equation}
Note that in this and all other cases, if we use the log-likelihood function as the basis for a minimization algorithm in fitting a model to data, we are comparing different values of the resulting log-likelihood for a particular parameter value with that resulting from another. Since this is done by calculating the difference between them, it implies that constant terms---those that do not involve the parameter---do not contribute, and can therefore be dropped from the expression of the log-likelihood, now used as the basis of a fit statistic.

\subsection{Normal Variables}

The univariate normal probability density function is
\begin{equation}
\label{eq:normalDensity}
g(x; \mu, \sigma) = \frac{1}{\sqrt{2\pi\sigma^2}} e^{-{(x-\mu)^2}/{2\sigma^2}},
\end{equation}
where the variable $x$ is a real number, $\mu$ is the mean, and $\sigma$ the standard deviation. There are now two parameters, $\mu$ and $\sigma$, instead of a single one as for the Poisson distribution. Because the two parameters are independent, (in practice only for a large enough number of measurements), the likelihood function can be expressed for either one of them (fixing the other) or for both, in which case, for a single measurement, it is:
\begin{equation}
\label{eq:normalLikelihood-single}
L(\mu,\sigma | x) \propto g(x; \mu, \sigma) = \frac{1}{\sqrt{2\pi\sigma^2}} e^{-{(x-\mu)^2}/{2\sigma^2}}
\end{equation}
and for a collection of measurements it becomes
\begin{equation}
\label{eq:normalLikelihood-joint}
L(\mu,\sigma | \pmb{x}) = \prod_i \frac{1}{\sqrt{2\pi\sigma^2}} e^{-{(x_i-\mu)^2}/{2\sigma^2}}.
\end{equation}
Note that we could drop the constant $1/\sqrt{2\pi}$ because the scale of the likelihood function can be normalized as we desire, but we keep it in order to retain the same form as the density function.  If we are comparing each of the $N$ measurements with the model-predicted values (we need to specify both $\mu_i$ and $\sigma_i$ for each values $x_i$), the likelihood function is
\begin{equation}
\label{eq:normalLikelihood-model}
L(\pmb{\mu,\sigma} | \pmb{x}) = \prod_i \frac{1}{\sqrt{2\pi{\sigma_i}^2}} e^{-{(x_i-\mu_i)^2}/{2\sigma_i^2}}
\end{equation}
and the log-likelihood is therefore given by
\begin{equation}
\label{eq:normalLogLikelihood-model}
\ln L(\pmb{\mu,\sigma} | \pmb{x}) = -\frac{1}{2} \left( N\ln(2\pi) +\sum_i \left[\ln(\sigma_i^2) + (x_i - \mu_i)^2/\sigma_i^2\right]\right).
\end{equation}

\subsection{Pearson Type III or $\chi^2$ Variables}

The $\chi^2$ probability density function is given by
\begin{equation}
f(x;k) = \frac{1}{2^{k/2}\Gamma(k/2)} x^{k/2 -1} e^{-x/2},
\end{equation}
where the variable $x$, is equal or greater than zero and continuous on the real line, and the parameter $k$, called the number of degrees of freedom (dof) of the distribution, is traditionally defined to be a positive integer, but can be any positive real number (how it is used here).  As is the case for the Poisson distribution, the parameter determines the shape of the curve.

The $\Gamma$ function is a generalized factorial defined as
\begin{equation}
\Gamma(x)=\int_0^\infty e^{-t} t^{x-1} dt
\end{equation}
and has the following relations:
\begin{equation}
\Gamma(x+1)= x\Gamma(x),\hspace{2mm} \Gamma(n) = (n-1)!\hspace{2mm}{\rm and}\hspace{2mm} \Gamma(1/2) = \sqrt{\pi}.
\end{equation}
The single-measurement likelihood function is therefore
\begin{equation}
\label{eq:chiSquaredLikelihood-single}
L(k|x) \propto f(x ; k) = \frac{1}{2^{k/2}\Gamma(k/2)} x^{k/2-1} e^{-x/2}
\end{equation}
and the joint likelihood for multiple measurements is
\begin{equation}
\label{eq:chiSquaredLikelihood-joint}
L(k|\pmb{x}) = \prod_i \frac{1}{2^{k/2}\Gamma(k/2)} x_i^{k/2-1} e^{-x_i/2}.
\end{equation}

For the general case of a collection of samples drawn from different $\chi^2$ distributions, each with a given value of $k$, the joint likelihood function is given by the generalization of Equation (\ref{eq:chiSquaredLikelihood-joint}) where $k$ is replaced by $k_i$:
\begin{equation}
L(\pmb{k} | \pmb{x}) = \prod_i \frac{1}{2^{k_i/2}\Gamma(k_i/2)} x_i^{k_i/2-1} e^{-x_i/2}.
\end{equation}
Taking the natural logarithm and simplifying the expression using the properties of the logarithm yields the log-likelihood:
\begin{equation}
\label{eq:chiSquareLogLikelihood}
\ln L(\pmb{k} | \pmb{x}) = \sum_i \left[ \left({\frac{k_i}{2}-1}\right)\ln x_i -\frac{x_i}{2} - \frac{k_i}{2}\ln 2 - \ln\Gamma\left(\frac{k_i}{2}\right) \right],
\end{equation}
where $\ln\Gamma(z)$ is available in most programming environments, but can also be approximated by
\begin{equation}
\ln\Gamma(z) \approx (z-1/2)\ln z - z + \frac{1}{2}\ln(2\pi).
\end{equation}
Substituting $z=k_i/2$ gives
\begin{equation}
	\ln\Gamma\left(\frac{k_i}{2}\right) \approx \frac{1}{2}(k_i-1)\ln\left(\frac{k_i}{2}\right) - \frac{k_i}{2} + \frac{1}{2}\ln(2\pi).
\end{equation}
In all equations, $k_i$ is the model-predicted numbers of dof in channel $i$, and $x_i$ is the value of the random variable measured in that channel. This implies that there is for each channel a different value of the parameter $k$ and a single measurement $x$. If there are several measurements for each channel, then each one would have a joint-likelihood function $L(k_i | \pmb{x})$ resulting from the product of $j$ measurements. The overall joint-likelihood would then be the product of these joint-likelihoods per channel.

The $\chi^2$ distribution is a single-parameter function: its mean is the number of dof, and the variance is twice that. This implies a large scatter about the mean that is proportional to it in amplitude. The mode---the most likely value where the probability peaks---is not the mean as for the normal and Poisson distributions; it is given by max($k-2$, 0).

\subsection{Exponential Variables}

The exponential probability density is given by
\begin{equation}
\label{eq:exponentialDensity}
f(x; \tau) = \frac{1}{\tau} e^{-x/\tau},
\end{equation}
where $x$ is the random variable---a real number greater or equal to zero, and $\tau$---a real number greater than zero, is called the decay constant because it defines the speed at which the function decays. Both the mean and standard deviation of the distribution equal $\tau$, the variance equals $\tau^2$, and the mode is always zero. 

Of particular relevance here is that inter-arrival times of a homogeneous Poisson process with mean rate $\mu$ are exponentially distributed according to $\tau=1/\mu$. So, naturally, $\mu=1/\tau$.

The single-measurement likelihood function is
\begin{equation}
\label{eq:exponentialLikelihood}
L(\tau | x) \propto f(x; \tau) = \frac{1}{\tau} e^{-x/\tau}.
\end{equation}
The joint-likelihood of a set of measurements drawn from the same distribution (same $\tau$, same frequency channel) is the product of the individual probabilities:
\begin{equation}
\label{eq:exponentialLikelihood-joint}
L(\tau) = \prod_i \frac{1}{\tau} e^{-x_i/\tau}.
\end{equation}
The joint-likelihood for a set of measurements---one pair $(x_i, \tau_i)$ per channel---is given by
\begin{equation}
\label{eq:exponentialLikelihood-model}
L(\pmb{\tau}) = \prod_i \frac{1}{\tau_i} e^{-x_i/\tau_i}.
\end{equation}
The corresponding log-likelihood function is given by
\begin{equation}
\label{eq:exponentialLogLikelihood-model}
\ln L(\pmb{\tau}) = -\sum_i (\ln\tau_i + x_i/\tau_i).
\end{equation}

\subsection{Inverse-exponential Variables}

The inverse-exponential density function is
\begin{equation}
\label{eq:inverseExponentialDensity}
f(x; t) = \frac{t}{x^2}  e^{-t/x},
\end{equation}
where $x$ is the random variable---a real number strictly greater than zero, and $t$---a real number greater or equal to zero, is the parameter of the distribution. Analogously to the exponential, both the mean and standard deviation are equal, but are given by the value of $2t$, the variance is therefore $4t^2$, and the mode, unlike the exponential, equals $t$/2. 

In relation to a homogeneous Poisson process with mean rate $\mu$, taking the inverse of the inter-arrival times to compute instantaneous rates will yield an inverse-exponential distribution with a mean of $2t=2\mu$. So, in this case, $\mu=t$, and the distribution's peak or mode is at $t/2$, therefore $\mu/2$.

The single-measurement likelihood function is
\begin{equation}
\label{eq:inverseExponentialLikelihood}
L(t | x) \propto f(x; t) = \frac{t}{x^2} e^{-t/x}.
\end{equation}
The joint-likelihood of a set of measurements drawn from the same distribution, as in all cases, is the product of the individual probabilities:
\begin{equation}
\label{eq:inverseExponentialLikelihood-joint}
L(t) = \prod_i \frac{t}{{x_i}^2} e^{-t/x_i}.
\end{equation}
The joint-likelihood for a set of measurements and model expectations---one pair $(x_i, t_i)$ per channel---is given by
\begin{equation}
\label{eq:inverseExponentialLikelihood-model}
L(\pmb{t}) = \prod_i \frac{t_i}{{x_i}^2} e^{-t_i/x_i}.
\end{equation}
And the corresponding log-likelihood function is given by
\begin{equation}
\label{eq:inverseExponentialLogLikelihood-model}
\ln L(\pmb{t}) = \sum_i (\ln t_i - 2\ln x_i - t_i/x_i).
\end{equation}

\bibliography{refs-final}

\end{document}